\documentclass[11pt]{article}

\usepackage[utf8]{inputenc} 
\usepackage[T1]{fontenc}
\usepackage{url}
\usepackage{ifthen}
\usepackage{cite}
\usepackage{graphicx}
\usepackage{algorithm}
\usepackage{algorithmic}
\usepackage{caption}
\usepackage{dsfont}
\usepackage{amsbsy}
\usepackage{amssymb}
\usepackage{amsmath}%
\usepackage{amsfonts}%
\usepackage{amsthm}
\usepackage{breqn}
\usepackage{graphicx}
\usepackage{colonequals}
\usepackage{psfrag}
\usepackage{mathrsfs}
\usepackage{bbm}
\usepackage{stfloats}
\usepackage{tikz}
\usepackage{pgfplots}
\usepackage{arxiv}
\usepackage{epstopdf}
\usepackage{subcaption}
\usepackage{array}

\newcommand{\blfootnote}[1]{%
  \begingroup
  \renewcommand{\thefootnote}{}
  \footnotetext{#1}%
  \addtocounter{footnote}{-1}
  \endgroup
}

\usepackage{xcolor}

\title{Optimized Couplings for Watermarking Large Language Models}
\date{}
\author{
Dor Tsur${}^*$\\
Ben Gurion University\\
\And
Carol Xuan Long${}^*$\\
Harvard University\\
\And
Claudio Mayrink Verdun\\
Harvard University\\
\And
Hsiang Hsu\\
JPMorganChase\\
\And
Haim Permuter\\
Ben Gurion University\\
\And
Flavio P. Calmon\\
Harvard University\\
}

\begin{document}

\maketitle


\begin{abstract} 
Large-language models (LLMs) are now able to produce text that is, in many cases, seemingly indistinguishable from human-generated content.
This has fueled the development of watermarks that imprint a ``signal'' in LLM-generated text with minimal perturbation of an LLM's output.
This paper provides an analysis of text watermarking in a one-shot setting.
Through the lens of hypothesis testing with side information, we formulate and analyze the fundamental trade-off between watermark detection power and distortion in generated textual quality.
We argue that a key component in watermark design is  generating a coupling between the side information shared with the watermark detector and a random partition of the LLM vocabulary.
Our analysis identifies the optimal coupling and randomization strategy under the worst-case LLM next-token distribution that satisfies a min-entropy constraint. 
We provide a closed-form expression of the resulting detection rate under the proposed scheme and quantify the cost in a max-min sense.
Finally, we provide an array of numerical results,comparing the proposed scheme with the theoretical optimum and existing scheme, in both synthetic data and LLM watermarking. Our code is available at \url{https://github.com/Carol-Long/CC_Watermark}
\end{abstract}

\section{Introduction}
\blfootnote{\textsuperscript{*}Equal contributions. } 
\blfootnote{\textsuperscript{a}Correspondence to: \texttt{carol\_long@g.harvard.edu, dortz@post.bgu.ac.il, flavio@seas.harvard.edu}}
\blfootnote{\textsuperscript{b}This paper was prepared by Hsiang Hsu prior to his employment at JPMorgan Chase $\&$ Co.. Therefore, this paper is not a product of the Research Department of JPMorgan Chase \& Co. or its affiliates. Neither JPMorgan Chase \& Co. nor any of its affiliates makes any explicit or implied representation or warranty and none of them accept any liability in connection with this paper, including, without limitation, with respect to the completeness, accuracy, or reliability of the information contained herein and the potential legal, compliance, tax, or accounting effects thereof. This document is not intended as investment research or investment advice or as a recommendation, offer, or solicitation for the purchase or sale of any security, financial instrument, financial product, or service or to be used in any way for evaluating the merits of participating in any transaction.} 
\blfootnote{\textsuperscript{c}This material is based upon work supported by the National Science Foundation under Grant No FAI 2040880, CIF 2231707, CIF 2312667. The work is also supported by a gift by J.P. Morgan.} 

\setcounter{footnote}{0}
A large language model (LLM) is a generative model that, given a string of input tokens, outputs a probability distribution $Q_X$ for the next token $X$ in the sequence. The emergence of LLMs that generate text that is largely indistinguishable from humans has led to the creation of trustworthy text generation algorithms \cite{huang2024trustllm} that create safe \cite{bai2022constitutional}, interpretable \cite{geva2021transformer}, and authentic \cite{lin2022truthfulqa} content.
This work focuses on \emph{watermarking}: the process of embedding a ``signal'' at the token level in LLM-generated text. The goal of a watermark is to enable automated detection of AI-generated content, providing proof of its authenticity (or lack thereof) and potentially of its origin. The past two years have witnessed the creation of increasingly sophisticated LLM watermarking schemes \cite{kirchenbauer2023watermark,christ2024undetectable,kuditipudi2023robust,zhaoprovable,aaronson2023watermark,he2024universally,bahri2024watermark,dathathri2024scalable,yang2023watermarking,ren2024subtle,huunbiased2024,zhao2024permute,chao2024watermarking,qu2024provably,xie2024debiasing,liuadaptive,fernandez2023three}. 

A hallmark of existing LLM watermarks is their reliance on either distorting or coupling the next-token distribution $Q_X$ with a random variable $S$ drawn from a known distribution $P_S$. Here, $S$ represents shared randomness known both by the watermark generator and detector and a coupling refers to a joint probability distribution $Q_{X,S}$ whose marginals are $Q_X$ and $P_S$, representing how the token distribution and side information are probabilistically related while preserving their individual distributions. For instance, \cite{kirchenbauer2023watermark} -- which ignited the recent interest in LLM watermarking in the machine learning community -- distorts $Q_X$ by randomly choosing a set of tokens (as determined by $S$) to be on a ``green list,'' i.e., a subset of tokens that are favored during generation,  and increasing the mass of those tokens accordingly. The detector then counts the number of tokens in a sequence that appears on the green list and declares the text watermarked (i.e., AI-generated) if this count exceeds a threshold. 
However, such a distortion of the LLM distribution may impair the textual quality.
Alternative approaches include \cite{aaronson2023watermark,kuditipudi2023robust,he2024universally,chao2024watermarking}, which instead couple $Q_X$ with the distribution $P_S$. Such couplings enable  ``distortion-free'' watermarks that (averaged over $P_S$) do not change the expected next-token distribution, yet are still detectable. 

The exact nature of the shared randomness $S$ between the model and the detector varies across watermark implementations. $S$ can be, for example, generated from the hash of previous tokens in a sequence \cite{kirchenbauer2023watermark} (where a hash function converts the token history into a fixed-size value that deterministically produces pseudo-random bits) or sophisticated tournament-like sampling strategies \cite{dathathri2024scalable}. 
For our theoretical analysis, we abstract away the exact generation process of the shared randomness $S$.

At a high level, existing LLM watermarks perform two steps when generating a sequence of tokens $\{X_i\}_{i=1}^n$ given shared randomness $\{S_i\}_{i=1}^n$:
\begin{enumerate}
    \item \emph{Watermark Generation:} For the $i$-th generated token and given $S_i$ and the predicted next token distribution $Q_{X}$, draw the next token by sampling from $X_i\sim Q_{X|S_i}.$ 
    \item \emph{Detection:} Given a sequence $\{(X_i,S_i)\}_{i=1}^n$, compute the statistic
        $T_n=\frac{1}{n}\sum_{i=1}^n f(X_i,S_i)$ for some function $f:\cX\times\cS\to[0,1]$,
    and declare that the sequence $\{X_i\}_{i=1}^n$ is watermarked if $T_n\geq \tau$.
\end{enumerate}
Importantly, a crucial assumption of current LLM watermarking schemes is that the function $f$ \underline{\emph{does not}} assume knowledge of the token distribution $Q_{X^n}$.  This allows watermarks that are directly detectable from the sequence $\{(X_i,S_i)\}_{i=1}^n$, i.e., directly from generated text, without accessing the underlying LLM. If the distribution of the generated tokens $Q_{X^n}$ was known, then a standard likelihood ratio test (LRT) would suffice for watermark detection. What makes LLM watermarking distinct from existing information-theoretic watermarking schemes (e.g., \cite{gel1980coding,willems2000informationtheoretical,chen2000design,moulin2003information,martinian2005authentication,villan2006text}) are the assumptions that (i) the source distribution is unknown to the watermark detector and (ii)  watermarking is performed on a per-token (vs. sequence) level.

\subsection{Main Contributions}\label{sec:main_contributions}
Motivated by the success of token-level schemes for LLM watermarking, we provide an in-depth analysis of a single-token watermarking process, i.e., when $n=1$.
Specifically, we study how to generate a coupling $Q_{X,S}$ and the corresponding detection function $f$ that maximizes the probability of detection of the watermark, while controlling the quality of the text.
The latter is controlled through the distortion relative to $Q_X$ -- a quantity we call \emph{perception}, following recent trends in the information theory literature on the source coding problem \cite{blau2019rethinking,theis2021coding,chen2022rate}. We refer to this setting as \emph{one-shot watermarking}. We jointly optimize $Q_{X,S}$ and $f$ given a perception constraint, with the case $\bar{Q}_X=Q_X$ corresponding to the \emph{perfect perception} setting. 
We focus on one-shot watermarking since, as mentioned above, existing schemes are constrained to watermark on a token-by-token basis. Moreover, small gains in single-token watermark detection compound to exponential gains in detection accuracy in threshold tests applied across multiple tokens.

We begin with an information-theoretic formulation for one-shot watermarking. We quantify the fundamental trade-off between watermark detection vs. perception when the underlying next-token distribution $Q_X$ is known with the side information $P_S$ uniformly distributed. This analysis yields a fundamental upper bound on one-shot watermark performance; see Theorems \ref{thm:opt_cornerpoints} and \ref{thm:universal_ub}. Interestingly, when the watermark does not change the next-token probability (i.e., perfect perception), optimizing a one-shot watermark is equivalent to maximizing the TV-information $ \mathsf{TV}\left(Q_{X,S}\|Q_XP_S\right)$ across the conditional distribution $Q_{X|S}$ -- a non-convex optimization problem \cite[Section~7]{polyanskiy2024information}. This formulation embeds TV-information with a new operational interpretation. 


We optimize one-shot watermarks when $Q_X$ is unknown to the detector but satisfies a min-entropy constraint, i.e., 
$\|Q_X\|_\infty \leq \lambda$ (Eq. \eqref{eq:maxmin}), which corresponds to $H_\infty(Q_X) \geq -\log(\lambda)$. Operationally, lower values of $\lambda$ correspond to higher entropy token distributions with greater uncertainty, while higher values of $\lambda$ indicate more concentrated distributions where the next token is more predictable.
Moreover, we optimize for detection tests of the form $\ones[f(X)=S]$,  where $f:\calX\to\cS$ forms a partition of $\cX$.  

Motivated by the fact that deterministic token partitions lead to low detection probabilities, we introduce randomness to $f$. In Theorem \ref{thm:optimal_maxmin_detection}, we characterize the probability of detection of such detection tests under the worst-case token distribution for any given min-entropy constraint and any size of side information.  
We pair our analysis with a characterization of the optimal design of the partition randomization.
In Theorem \ref{thm:approx_maxmin_detection}, we consider a simplified token partition strategy and show that it yields a near-optimal detection probability.
Together, we provide the complete characterization of the minimax detection rate for a given vocabulary size, side information, and min-entropy constraint under the optimal and near-optimal partition randomization strategies, plotted in Figure \ref{fig:Rd_star}.

We provide an array of numerical results of the Correlated Channel (CC).
On synthetic data, we show how the CC watermark detection meets the theoretical optimum, while outperforming the red-green watermark \cite{kirchenbauer2023watermark}. We investigate the effect of the side information alphabet and demonstrate performance in a sequential setting. Finally, we provide empirical results on LLM watermarking using the Waterbench dataset \cite{tu2023waterbench}.



\textbf{Related Work.} 
Watermarking has been extensively studied in information theory \cite{chen2000design,moulin2003information,martinian2005authentication}, particularly through the Gelfand-Pinsker (GP) channel \cite{gel1980coding,villan2006text,willems2000informationtheoretical}.
These approaches typically focus on watermarking sequences via joint typicality and assume perfect knowledge of the underlying source distribution. 
The  work of \cite{kirchenbauer2023watermark} led to various developments in watermarking schemes \cite{aaronson2023watermark,he2024universally, bahri2024watermark,dathathri2024scalable,yang2023watermarking,ren2024subtle,huunbiased2024,zhao2024permute,chao2024watermarking,qu2024provably,xie2024debiasing,liuadaptive}, with several approaches that focus on distortion-free methods, e.g., \cite{kuditipudi2023robust,huunbiased2024,zhao2024permute,christ2024undetectable}. In particular, \cite{chao2024watermarking} proposes a watermark using error-correcting codes leading to correlated channels similar to the ones we find via optimizing couplings. 
In \cite{huang2023towards}, the optimal Type-II error for the bounded Type-I error is analyzed by comparing watermarking schemes with the uniformly most powerful watermark with knowledge of $Q_X$. 
The authors of \cite{he2024universally} characterize the universal Type II error while controlling the worst-case Type-I error by optimizing the watermarking scheme and detector.
While these works operate on a token-level basis, they focus on the effect of a given strategy along a sequence.
In contrast, we focus on a preliminary step and aim to answer the simple yet important question -- \textit{What is the optimal coupling when watermarking a single token?}\footnote{Our code is available at \url{https://github.com/Carol-Long/CC_Watermark}}

\begin{figure}[!tb]
    \centering
    \includegraphics[trim={10pt 70pt 13pt 40pt}, clip, width=0.8\linewidth]{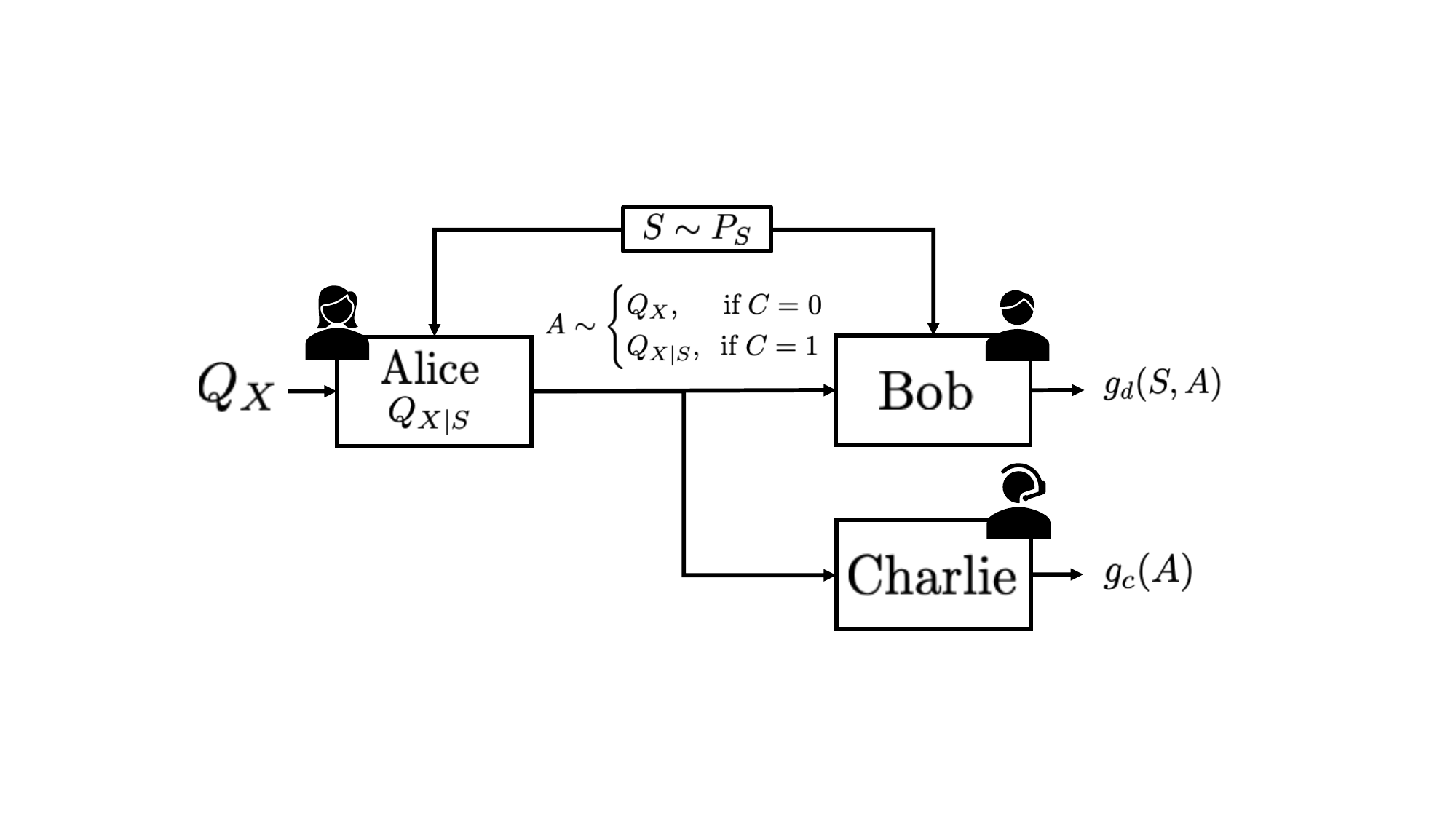}
    \caption{Watermarking problem as a hypothesis test with side information.}
    \label{fig:operational_problem}
\end{figure}

\section{Optimal One-Shot Watermarking}
In this section, we formulate the watermarking problem, derive the resulting optimization problem, and discuss the optimal solution structure.
We focus on the fundamental trade-off between detection probability and perceptual quality.
As mentioned above, while the optimal approach to watermarking considers sequence-to-sequence schemes, due to the autoregressive nature of token generation in LLMs most popular schemes focus on token-level strategies \cite{kirchenbauer2023watermark,aaronson2023watermark,he2024universally,dathathri2024scalable}.
As a first step towards token-level watermarking of sequences, we provide an extensive analysis of the one-shot setting.
We discuss the extension to a token-level scheme in the sequential case in Section \ref{sec:sequential}.

\subsection{Problem Setting}
We consider a hypothesis test using the private side information setting and textual quality of the model as the ability of an external observer to detect the watermark without access to the side information. Formally, let $Q_X$ be the LLM distribution over some finite vocabulary of $|\cX|=m$ tokens. 
We consider \textit{Alice} (the watermarker), whose goal is to convey a single token to \textit{Bob} (the detector), which, in turn, tries to detect whether the token is watermarked or not.
Alice and Bob share some random side information\footnote{Side information often corresponds to a secret shared key; see, e.g., \cite{kuditipudi2023robust, zhao2024sok}.} $S\sim P_S$ with $|\cS|=k$.
Furthermore, we consider \textit{Charlie} (average observer), which tries to detect the existence of the watermark but does not have access to the side information.
The setting is depicted in Figure \ref{fig:operational_problem}.

On Alice's end, the watermark design boils down to the construction of the conditional distribution $Q_{X|S}$. We consider a Bayesian setting, in which Alice transmits a token according to the outcome of a fair coin $C$. The transmitted token is thus given by: 
\begin{equation}\label{eq:sample_a}
    A = 
    \begin{cases}
        X\sim Q_X &\mbox{if }C=0,\\
        \tilde{X}\sim Q_{X|S} &\mbox{if }C=1.
    \end{cases}, \quad C\sim\mathsf{Ber}\left(\frac{1}{2}\right)
\end{equation}
where $C\indep(X,\tilde{X},S)$.
To detect the watermark, Bob performs the following hypothesis test
\begin{align*}
&H_0: A\sim Q_X\\
&H_1: A\sim Q_{X|S}.
\end{align*}
We assume that Charlie is aware of the watermarking mechanism but is not aware of the specific sample of $S$. Therefore, Charlie performs an hypothesis test with a corresponding alternative hypothesis, i.e.
\begin{align*}
&H_0: A\sim Q_X\\
&H_1: A\sim\bar{Q}_X,
\end{align*}
where $\bar{Q}_X\triangleq\EE_{S}[ Q_{X|S}]$ is the watermark distribution averaged w.r.t. the side information $S$, while $QX|S$ is the conditional token distribution given specific side information. Note that $\bar{Q}_X$ is distinct from $Q_{X,S}$, which would denote their joint distribution.

\subsection{A Detection-Perception Perspective}
Given the hypothesis test formulation, we recast the problem of watermarking as a trade-off between two measures: Bob's \textit{detection} and Charlie's \textit{perception} probabilities. 
Motivated by recent advances in lossy source-coding \cite{blau2019rethinking,theis2021coding,chen2022rate}, we adopt the notion of perceptual qualities of the data, which is quantified through a discrepancy measure between the two distributions, e.g. $f$-divergences, rather than a metric calculated directly on the random variables.

We define two fundamental metrics that capture the trade-off between detection capability for Bob and imperceptibility for Charlie. For Bob's detection capability, we weigh true negative (TN) detections with prior $\pi_0$ and true positive (TP) detections with prior $\pi_1=1-\pi_0$. The tests are defined as follows:
\begin{definition}[Watermark Tests and Error Probabilities]
    A watermarking scheme comprises of a detection test  $g_d: \mathcal{X} \times \mathcal{S} \to \{0,1\}$, such that for $(A,S)\in\cX\times\cS$, we respectively define the detection probability with prior $\pi=(\pi_0,\pi_1)$ as
    \begin{align*}
    \rd \triangleq \EE_{\pi}\left[\Pr(g_d(S,A)=C)\right],
    \end{align*}
    where $C \in \{0,1\}$ is the label indicating whether the token is watermarked and the expectation is with respect to all randomness in the system, i.e., the side information S and token distributions. Perception probability $\rp$ is similarly defined with a test $g_c: \mathcal{X} \to \{0,1\}$ and a uniform prior $\pi_0=1/2$.
\end{definition}


\noindent Optimally, we aim to maximize detection $\rd$ while minimizing $\rp$, which indicate Charlie's low perception of the watermark. Next, we formalize the trade-off.

\subsection{Characterizing Optimal Trade-off}
Following the Neyman-Pearson Lemma \cite{lehmann1986testing}, the likelihood ratio gives the optimal test statistic, resulting in a simple closed form for ($\rd$,$\rp$) in terms of $\mathsf{E}_\gamma$ (or hockey-stick) divergence. The next proposition is a direct result of the well-known connection between $\mathsf{E}_\gamma$ and hypothesis testing; see, e.g., \cite{polyanskiy2010channel,polyanskiy2010channel_,liu2016e}.

\begin{prop}\label{prop:tv_prob}
Fix $(P_S,Q_X, Q_{X|S})$ and error priors $\pi_0$ and $\pi_1$. Let $\gamma = \frac{\pi_1}{\pi_0}$. Using the LRT, the optimal detection and perception probabilities are given by 
\begin{align}
\rd &= \pi_1 + \pi_0 \mathsf{E}_\gamma\left( Q_{X|S}P_S, Q_XP_S\right), \\
    \rp &= \frac{1}{2} + \frac{1}{2}\TV\left(\bar{Q}_X, Q_X\right).
\end{align}
\end{prop}

\begin{remark} The $\mathsf{E}_\gamma$-divergence characterizes the error of hypothesis tests with specified priors on TP and TN rates. 
It can be defined as\footnote{Some works include a residual term $(1-\gamma)_+$ in the $\mathsf{E}_\gamma$-divergence definition, e.g., \cite{asoodeh2020contraction}, which we omit as it does not affect the optimization problem.} \cite{liu2016e}
$$\mathsf{E}_{\gamma}(P,Q) \triangleq \max_{\calA} [P(\calA)-\gamma Q(\calA)],$$ where $\cA$ are rejection regions, $P(\calA)$ and $Q(\calA)$ are $1-$TN rate and TP rate, respectively.
When. $\pi_0=\pi_1$ and $\gamma = 1$, detection probability boils down to the total variation (TV) distance, in which case, we have $\rd = \frac{1}{2} + \frac{1}{2}\TV( Q_{X|S},Q_{X}|P_S)$, where $\TV( Q_{X|S},Q_{X}|P_S)=\TV( Q_{X|S}P_S,Q_{X}P_S)$ is termed the TV-information between $X$ and $S$.
\end{remark}

Our hypothesis testing framework employs priors $\pi_0$ and $\pi_1$ to explicitly weight the importance of different error types in the detection process.  Setting $\pi_0 = \pi_1 = \frac{1}{2}$ gives equal importance to both errors, whereas asymmetric values prioritize either minimizing false positives (incorrectly flagging human content as AI-generated) or false negatives (failing to detect AI-generated content). This Bayesian framework provides a principled approach to designing watermark schemes with detection rates optimized for specific operational requirements, where the relative costs of different error types may vary significantly across applications.
Due to Jensen's inequality, for any fixed $(P_S, Q_{X|S})$, we have $\rp \leq \rd$, i.e., Bob's access to the shared side information allows for a potentially higher detection probability.

It is important to note that our approach fundamentally relies on the concept of probability couplings, which is formally defined as a joint distribution $Q_{X,S}$ whose marginals are $Q_X$ and $P_S$. In our watermarking framework, once $P_S$ is fixed, finding the optimal coupling $Q_{X,S}$ is equivalent to optimizing the conditional distribution $Q_{X|S}$, since $Q_{X,S}(x,s) = Q_{X|S}(x|s)P_S(s)$. This approach allows us to construct ``distortion-free'' watermarks that maintain the expected token distribution while enabling detection.

Generally, for any perception constraint $\alpha_p\in[1/2,1]$, the optimal detection probability is given by the solution to the following optimization problem:
\begin{equation}\label{eq:curve_opt_iid}
    \sup_{ Q_{X|S}}{\mathsf{E}_\gamma\left( Q_{X|S}, Q_X|P_S\right)},\quad \textrm{s.t.}\quad  \TV\left(\bar{Q}_X,Q_{X}\right)\leq \alpha_p.
\end{equation}
We are interested in characterizing the $(\rd,\rp)$ trade-off region, which amounts to solving \eqref{eq:curve_opt_iid} as a function of $\alpha_p$.

Note that \eqref{eq:curve_opt_iid} is a non-convex optimization problem, which generally lacks a closed-form solution.
In what follows, we characterize the several corner points of the optimal curve (i.e., $\rp=0.5$), which, in turn, gives insight into the structure of the $(\rd,\rp)$ region within the box $[\frac{1}{2},1]^2$.
 
We begin by characterizing upper and lower bounds when $\alpha_p=0$, i.e., under \textit{zero-perception}. In this case, $Q_X=\bar{Q}_X$, which implies perfect textual quality. The following result establishes tight bounds on the optimal detection probability in this regime
\begin{thm}[Zero perception bounds]\label{thm:opt_cornerpoints}
     Fix $Q_X$ and let $P_S$ be uniform over $\cS$, with  $|\cS|\leq|\cX|$ and let $\pi_1=\frac{1}{2}$. Then, for $\rp=\frac{1}{2}$, we have
    \begin{equation}
    \label{eq:tv_opt_both}
    \frac{1}{2} \leq \sup_{ Q_{X|S}}\rd \leq \max\left(\frac{1}{2},1-\frac{\gamma}{2k}\right).
    \end{equation}
\end{thm}
\noindent The upper bound emerges from jointly optimizing over both the coupling $ Q_{X|S}$ and $Q_X$. This optimization reduces to a convex problem over the probability simplex, which we recast as counting the optimally assigning elements of $\cX$. The lower bound is achieved when $Q_X$ is a singleton.


Beyond bounding detection under zero-perception, we derive an upper bound on the detection probability that holds across all perception levels.
The bound is given as follows: 
\begin{thm}[Uniform detection upper bound]\label{thm:universal_ub}
Let $Q_{\mathsf{min}}\triangleq\min_{x\in\cX}Q_X(x)$. For any $\rp\geq 0$ we have $\rd\leq1-\frac{\gamma Q_{\mathsf{min}}}{2}$.
\end{thm}
This bound emerges from analyzing a simple strategy of replacing each token with on of the least probable elements in $\cX$,. i.e., some $x\in\cX$ that is attained with probability $Q_{\mathsf{min}}$.
The non-convexity of \eqref{eq:curve_opt_iid} is demonstrated in our experimental results, see Section \ref{sec:exp}, where exact solvers are used to compute the trade-off region.
In light of this challenge, we will next derive a simple and tractable watermarking scheme.

\section{A One-Shot Watermarking Scheme}\label{sec:cc_scheme}
While the optimal test that maximizes Bob's detection accuracy is the LRT, it is infeasible in practical scenarios where Bob is not assumed to have access to $Q_X$.
To make use of the shared side information, Bob and Alice look for a mechanism that couples $S$ with the token distribution.
This can be done by applying a map $f:\cX\to\cS$. 
Alice uses $(f(X),S)$ to construct a watermarked distribution, and Bob uses $(f(A),S)$ to detect its presence.
We note that a map $f$ creates a partition of $\cX$ into $\cS$ bins.
When $k=2$, this can be interpreted as a partition of $\cX$ into a rejection region and its complement.
We note that considering deterministic mappings is insufficient, as for $S\sim\mathsf{Unif}([1:k])$, the detection probability is $\frac{1}{k}$, independent of the choice of $(f,Q_X)$. Therefore, we introduce randomness into our partitioning approach by making the function $f$ stochastic rather than deterministic. Specifically, we define a randomized mapping that varies the way tokens are assigned to each partition based on additional random variables that both Alice and Bob can access.

\subsection{Optimal Randomized Partition -- Correlated Channel}
We randomize $f$ by introducing a set of $m$ $\cS$-valued random variables denoted $B^m$.
We assume that $B^m$ is publicly available to all parties and is therefore not considered a part of the private side information $S$.
Our goal is therefore to couple the side information with the randomized mapping $f(X,B^m)$.
This boils down to finding a coupling of $Q_X$ and $S$ through the design of partition randomness $P_{B^m}$ and conditional distribution $ Q_{X|S}$.
We look for such $(P_{B^m}. Q_{X|S})$ that are optimal under the worst choice of token distribution $Q_X$ within a given class.
Our problem is therefore formally given by the following max-min expression
\begin{align}\label{eq:maxmin}
    \rd^\star(\lambda) &\triangleq \max_{P_{B^m}} \min_{\substack{Q_X \in \Delta_m\\ \|Q_X\|_\infty \leq \lambda}} \mathbb{E}\left[\rd(Q_X,B^m)\right],
\end{align}
where $\|Q_X\|_\infty = \max_{x\in\cX}Q(x)$ and $\rd(Q_X,B^m)$ denotes the detection probability for a given token distribution $Q_X$ and partition randomness $B^m$.. As discussed in Section \ref{sec:main_contributions}, we consider the constraint $\{Q_X \in \Delta_m, \|Q_X\|_\infty \leq \lambda\}$ which enables a more comprehensive analysis by allowing us to adjust the parameter $\lambda$. This flexibility provides insights across various scenarios: smaller $\lambda$ values yield higher entropy token distributions with greater uncertainty, while larger $\lambda$ values produce more deterministic distributions with reduced uncertainty about the next token.

According to \eqref{eq:maxmin}, given a fixed pair $(P_{B^m},Q_X)$, we maximize $R_d(Q_X,B^m)$ by designing the coupling of $(f(X,B^m),S)$.
We consider the mapping of the form\footnote{We consider a vocabulary $\cX=[1:m]$, which can be thought of as the enumeration of the tokens.} $f(x,b^m)=b_x$ under which, the partition's probabilities are characterized by the distribution of the random variable $ Y\triangleq f(X,B^m)$.
To this end, we first solve the following optimization problem:
\begin{equation}\label{eq:ot_wm}
    \sup_{P_{S, Y}}\Pr(S= Y),\quad S\sim\mathsf{Unif}\left(\cS\right),  Y\sim P_{ Y}.
\end{equation}
This is a maximum coupling problem whose closed-form solution is given below. It is a direct consequence of the inf-representation of TV distance \cite{polyanskiy2022ITbook}.

\begin{prop}\label{prop:optimal_coupling}
Let $S\sim\mathsf{Unif}[1:k]$ and $P_{ Y}=\{p_1,\dots,p_k\}\in\Delta_k$, $t = \TV(P_S, Y)$ and let $\Xi$ be the set of all couplings of $(P_S,P_{ Y})$.
Then, $\argmax_{\pi \in \Pi}\Pr(S= Y)$ is given by 
$$
\pi( Y=i, S=j) = 
\begin{cases}
    \min(\frac{1}{k},p_i),\qquad i=j,\\
    \frac{1}{t}(\frac{1}{k}-p_i)(p_j - \frac{1}{k}),\: (i\in A) \cap (j\in A^c),\\
    0, \qquad \text{otherwise},
\end{cases}
$$
where $A = \{i: p_i \geq \frac{1}{k}\}$, and $A^c = [k] \setminus A$.
\end{prop}
The resulting coupling can be thought of as a transition kernel that maps $P_{Y}$ to $P_S$ under maximum acceptance probability. When $k=2$, the optimal coupling boils down to a  binary asymmetric channel, known in information theory as the Z-channel \cite{CovThom06}.
That is, when $S=0$, the mapping always outputs $ Y=0$, but when $S=1$, the mapping may output either $ Y=1$ or $ Y=0$ with certain probabilities. This asymmetric structure is particularly effective for watermark detection because it creates a distinctive pattern that appears only in watermarked content. We therefore term this method as the correlated channel (CC) watermark.
We note that CC was previously considered, for example, in \cite{chao2024watermarking}.

\begin{algorithm}[!tb]
\caption{Correlated Channel Watermark (CC)}
\label{alg:correlated_Channel}
\begin{algorithmic}[1]
\REQUIRE LLM distribution $Q_X$, Side information $S$, shared randomness $B^m$.
\STATE \textbf{Alice:} 
\STATE Generate $Q_{X|S,B^m}$ according to \eqref{eq:wm_dist_channel}
\STATE Flip a coin $C\sim\mathsf{Ber}(\frac{1}{2})$ and sample $A$ according to \eqref{eq:sample_a}.
\STATE \textbf{Bob:}
\STATE \textbf{if} {\( S=f(A,B^m) \)} --  Declare \textbf{Watermarked}
\STATE \textbf{else} -- Declare \textbf{Not watermarked}
\end{algorithmic}
\end{algorithm}

The CC scheme consists of the following steps: Both Alice and Bob observe $(s,b^m)$. Alice samples $C\sim\mathsf{Ber}(\frac{1}{2})$.
If $C=0$, she samples $a\sim Q_X$ and sends it. Otherwise, she samples and sends $a\sim Q_{X|S=s}$, which is given by the CC:
\begin{equation}\label{eq:wm_dist_channel}
    Q_{X|s,b^m}(x) = Q_X(x)\frac{P_{S| Y}(s|f(x,b^m))}{P_S(s)}.
\end{equation}
Bob performs the detection test by declaring that $a$ is watermarked if $s=f(a,b^m)$.
The complete list of steps is summarized in Algorithm \ref{alg:correlated_Channel}.
Note that by coupling $(P_{ Y},P_S)$, we result with a coupling of $(Q_X,P_S)$.
Consequently, we have
$Q_X = \EE_{S}[ Q_{X|S}]=\bar{Q}_X$, which implies that the CC watermark has zero perception.

\subsection{Theoretical Analysis of the CC scheme}
Given the optimal coupling, we give a closed-form expression for $\rd$ in terms of the TV surrogate of mutual information in the resulting channel.
\begin{prop}\label{thm:detection_close_form}
    The CC watermark detection is given by 
    \begin{equation}
    \label{eq:Rd_cc}
        \rd = \frac{1}{2}\left(1+\TV\left(P_{S},P_{S| Y}|P_{ Y}\right)\right) 
        = 1-\frac{1}{2k}-\frac{1}{2}\mathsf{TV}\left(P_{ Y},P_S\right).
    \end{equation}
\end{prop}
Proposition \ref{thm:detection_close_form} provides a closed-form characterization of Bob's detection probability as a function. Specifically, for $k=2$, we have $\rd = \frac{1}{2}(1+\tilde{p})$, where $\tilde{p}\triangleq\min\left(\tilde{p}_0,\tilde{p}_1\right)$.
This term is maximized when $ Y\sim\mathsf{Ber}(\frac{1}{2})$, with maximum value of $\frac{3}{4}$.
A consequence of Proposition \ref{thm:detection_close_form} is that we are interested in designing a partition that is as close as possible to $P_S$ as possible.
As $P_S$ is uniform over $\{1,\dots,k\}$, our aim is to obtain a uniform distribution, i.e., a balanced partition of the token vocabulary $\cX$, given the token distribution $Q_X$ and the partition randomness $P_{B^m}$.
\begin{remark}[Equivalence to the likelihood ratio test] When we consider the indicator test $\mathbf{1}\{f(x,b^m)=s\}$, the decision region obtained by the CC watermark is equivalent to the one attained by the LRT with threshold value of $\tau=1$. This follows from the observation that $\Pr[S|f(S,B^m)]\geq \frac{1}{2}$, if and only if $S=f(X,B^m)$. \label{thm:detection_test_optimal} 
\end{remark}

Next, we discuss the design of randomness. Specifically, we analyze the dependence of the CC watermark detection probability on the distribution of $B^m$ and propose an optimal design of $P_{B^m}$.


\subsection{Optimizing the Partition}\label{sec:bm_design}

As seen in Equation \eqref{eq:Rd_cc}, the distribution of the resulting partition governs the detection power of the CC watermark.
The partition distribution is determined by the token distribution $Q_X$ and the distribution of $B^m$.
As $Q_X$ cannot be controlled by the designer, we aim to characterize the class of distributions $P_{B^m}$ that maximizes $\rd$ under the worst-case adversarial distribution $Q_X$.
Due to the symmetry of the CC, we can restrict the optimization over permutation classes of $P_{B^m}$.
First, we show that the optimal $P_{B^m}$ is permutation invariant. 

\begin{lem}
\label{lem:permutation_invariance}
Let $F(P_{B^m})\triangleq \min_{\substack{Q_X\in \Delta_m\\ \|Q_X\|_\infty \leq \lambda}} \mathbb{E}_{P^\star_{B^m}}\left[\rd(Q_X,B^m)\right]$. Let $P^\star_{B^m}$ be a distribution that maximizes $F(P_{B^m})$. Consider a permutation: $\phi: \calS^m \to \calS^m$. Define $\tilde{P}_\phi(B^m) = P^\star_{B^m}(\phi \circ B^m)$ Then, $F(P_{B^m}) = F(\tilde{P}_{\phi})$.
\end{lem}

Next, let $\calP_m = \{\calB_1, ..., \calB_K\}$ be the partition of $\calS^m$ into sets of sequences that are identical up to a permutation, with $|\calP_m|=K$. We refer to each $\calB_i$ as a permutation class. We proceed to characterize the optimal mean detection probability $\rd^\star$ and the corresponding distribution $P^\star_{B^m}$.  

\begin{figure}[!t]
    \centering
    \begin{minipage}{0.2\textwidth}
    \centering
    \begin{tikzpicture}[scale=0.8]
      \node at (-1.1,1) {\small$ Y$};
      \node at (0,1.5) {\small$0$};
      \node at (0,0.5) {\small$1$};
      \node at (-0.6,1.5) {\small$\tilde{p}_0$};
      \node at (-0.6,0.5) {\small$\tilde{p}_1$};
      \node at (3.2,1) {\small$S$};
      \node at (2.5,1.5) {\small$0$};
      \node at (2.5,0.5) {\small$1$};
      \node at (2.9,1.5) {\small$\frac{1}{2}$};
      \node at (2.9,0.5) {\small$\frac{1}{2}$};
    
      \draw[->] (0.2,1.5) -- (2.2,1.5);  
      \draw[->] (0.2,1.5) -- (2.2,0.5);  
      \draw[->] (0.2,0.5) -- (2.2,0.5);  
      \node at (1,1.8) {\scriptsize$1-\beta(\tilde{p}_0)$}; 
      \node at (1.55,1.2) {\scriptsize$\beta(\tilde{p}_0)$}; 
      \node at (0.9,0.7) {\scriptsize$1$}; 
\end{tikzpicture}
    \end{minipage}
    \hspace{0.4cm}
    \begin{minipage}{0.2\textwidth}
    \centering
    \begin{tikzpicture}[scale=0.8]
      \node at (-1.1,1) { \small$ Y$};
      \node at (0,1.5) {\small$0$};
      \node at (0,0.5) {\small$1$};
      \node at (-0.6,1.5) {\small$\tilde{p}_0$};
      \node at (-0.6,0.5) {\small$\tilde{p}_0$};
      \node at (3.2,1) {\small$S$};
      \node at (2.5,1.5) {\small$0$};
      \node at (2.5,0.5) {\small$1$};
      \node at (2.9,1.5) {\small$\frac{1}{2}$};
      \node at (2.9,0.5) {\small$\frac{1}{2}$};
    
      \draw[->] (0.2,1.5) -- (2.2,1.5);  
      \draw[->] (0.2,0.5) -- (2.2,1.5);  
      \draw[->] (0.2,0.5) -- (2.2,0.5);  
      \node at (1,1.8) {\scriptsize$1$}; 
      \node at (0.7,1.2) {\scriptsize$\beta(\tilde{p}_1)$}; 
      \node at (1.7,0.71) {\scriptsize$1-\beta(\tilde{p}_1)$}; 
    \end{tikzpicture}
    \end{minipage}
    \caption{Optimal coupling between side information $S$ and random partition $ Y=f(X,B^m)$ for \( \tilde{p}_1 \leq 0.5 \) (\textbf{left}), \( \tilde{p}_0 \leq 0.5 \) (\textbf{right}), with $\beta(p)=\frac{2p-1}{2p}$.}
    \label{fig:Z-channel}
\end{figure}
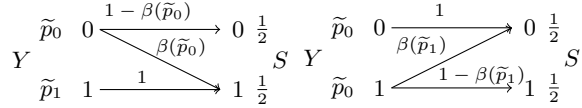

\begin{thm}[Optimal max-min Detection]
\label{thm:optimal_maxmin_detection}
Let $|\calS|=k$ and $\calX = m$, and assume that $m$ is divisible by $k$. Given min-entropy constraint $\lambda \in [0,1]$, and let $t = \left\lfloor \frac{1}{\lambda} \right\rfloor$. The optimal minimax detection probability from Equation \ref{eq:maxmin} is given by:
\begin{align}
    R_d^*(\lambda)  = 1-\frac{1}{2k}-\frac{1}{4}\EE[g(Q_\lambda^*,B^m)],
\end{align}
where 
\begin{align*}
    \EE[g(Q_\lambda^*,B^m)]=k 
    \sum_{c=0}^t \frac{\binom{m/k}{c}\binom{m-m/k}{t-c}}{\binom{m}{t}}
    &\left( \left(\frac{ (m/k)-c}{m-t} \right)\left| c\lambda +(1-\lambda t) -\frac{1}{k} \right|\right. \\
    &+ \left. \left(1-\frac{ (m/k)-c}{m-t} \right)\left| c\lambda -\frac{1}{k} \right|\right).
\end{align*}
Furthermore, the optimal detection probability is achieved for $P^*_{B^m}$ corresponding to uniform sampling over the permutation class of the sequence with an equal number of each element. For $|\calS|=2$, $P_{B^m} = \mathsf{Unif}(B^\star)$, where $B^\star = \{b^m\in\{0,1\}^m|b^m\text{ has equal number of 1's and 0's}\}$. 
\end{thm}
Under additional assumptions, we can further simplify the optimal detection.
\begin{cor}
    Under the setting of Theorem \ref{thm:optimal_maxmin_detection}, assume that $\lambda = \frac{1}{k}$. Then, we have 
\begin{align}
    R_d^*(\lambda)  = 1-\frac{1}{2k} - \frac{1}{2}\frac{\binom{(k-1)m/k}{k}}{\binom{m}{k}}.
\end{align}
Furthermore, if $k=2$ and $\lambda \in [\frac{1}{3},1]$ we have
\begin{equation}\label{eq:opt_b_thm}
    \rd^\star(\lambda) = 
    \begin{cases}
        \displaystyle\frac{3}{4}-\frac{m\lambda -1}{4(m-1)}, \quad~\mbox{for}\quad ~\frac{1}{2}\leq \lambda \leq 1\\
        \displaystyle\frac{3}{4} - \frac{m-2}{8(m-1)}, \quad~\mbox{for}\quad ~\frac{1}{3}\leq \lambda <\frac{1}{2}.
    \end{cases} 
\end{equation}
\end{cor}

Here, we have characterized detection for the worst-case distributions $Q^\star_\lambda$, which lie at the extreme point of the feasible set --- probabilities with bounded inf norm $\|Q_X\|_\infty \leq \lambda$). For example, for $\lambda \in [0.5,1]$, the above minimax detection probability corresponds to token distributions with only two nonzero entries, i.e., $Q_X$ takes the form $[\lambda,1-\lambda,0,...,0]$;
for $\lambda \in [\frac{1}{3},\frac{1}{2}]$, the worst-case token distribution have 3 non-zero elements and has the form $[\lambda,\lambda, 1-2\lambda,0,...,0]$.
Furthermore, we note that due to Equation \eqref{eq:Rd_cc}, when $k=2$, $\rd$ is upper bounded by $\frac{3}{4}$.
Thus, the second term in \eqref{eq:opt_b_thm} serves as a penalty when considering the max-min setting.
Notably, for $\lambda \in [0.5,1]$ and when $m$ is large, this penalty equals $\frac{\lambda}{4}$, which implies that the cost of considering worst-case token distributions is lower bounded by $\frac{1}{8}$.

\begin{figure}[tb]
    \centering
    \includegraphics[width=0.45\linewidth]{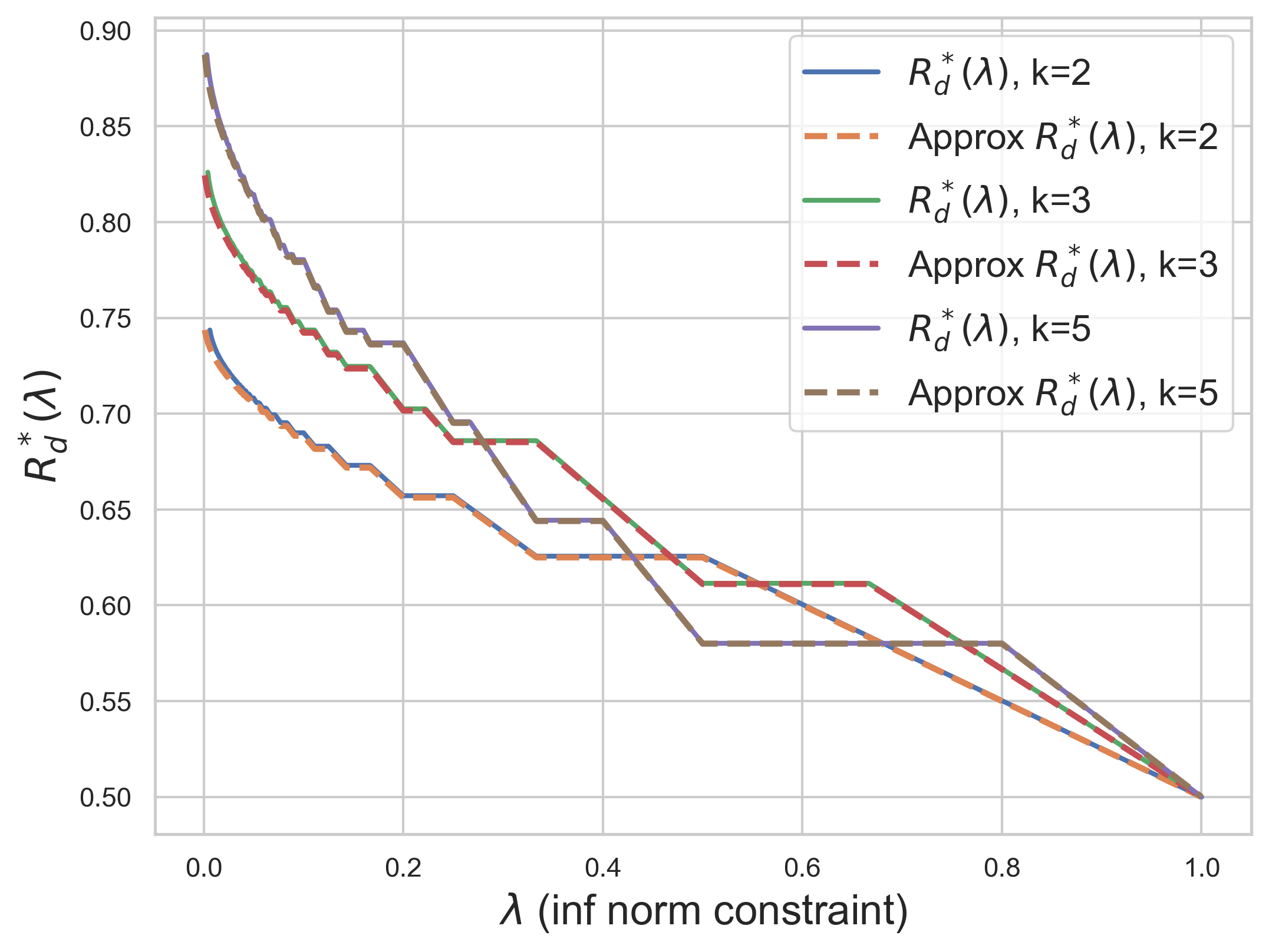}
    \caption{Optimal detection probability of CC in one-shot on the adversarial token distribution (Eq. \ref{eq:maxmin}) is plotted against the inf-norm constraint $\lambda$ (or equivalently, an entropy constraint) on $Q_X$$^3$.
    When $\lambda = 1$ (entropy $H(Q_X)=0$), $Q_X$ is deterministic, and detection is random. As entropy of $Q_X$ grows (moves to smaller $\lambda$ values), single-token optimal detection probability reaches a maximum of around 0.75 for binary side information. If the side information one transmits contain a larger set of values, CC achieves a higher detection probability correspondingly. The actual detection rate (solid lines) and approximate solutions (dotted lines) overlap for large enough vocabulary size$^4$, and their exact forms are provided in Theorem \ref{thm:optimal_maxmin_detection} and \ref{thm:approx_maxmin_detection}.}
    \label{fig:Rd_star}
\end{figure}

In addition to characterizing the minimax detection rate, Theorem \ref{thm:optimal_maxmin_detection} shows that the optimal sampling strategy for token partition $B^m$ is to sample uniformly from a collection of sets with an equal number of each element in $k$. 
Next, we show that we can adopt a much simpler sampling strategy, sampling i.i.d. Bernoulli variables with probability $\frac{1}{k}$ and arrive at a near-optimal detection probability.
In Figure \ref{fig:Rd_star}, we plot the probability of detection of both sampling strategies and show that the Bernoulli sampling strategy results in negligible approximation error.
To motivate i.i.d. Bernoulli sampling, we start with an alternative view of the optimal sampling strategy in Theorem \ref{thm:optimal_maxmin_detection}. Sampling a $b^m$ uniformly over $\calB^*$ --- containing sequences with equal numbers of each element in $k$ --- can be equivalently defined as the following process: given $m$ elements with predefined proportions $[\frac{1}{k},...,\frac{1}{k}]$, sample $m$ times with replacement.
In the following theorem, we obtain an approximation of $R^\star_d$ for any $\lambda$ by sampling without replacement. We also show that, by applying de Finetti's theorem on finite exchangeable sequences\cite{diaconis1980finite}, the approximation error decays with $O(\frac{1}{m})$.

\footnotetext{For discrete probability $Q_X$, inf-norm and entropy are connected via $H(Q_X)\geq -\log\|Q_X\|_{\infty}$, and we have $\lambda = \|Q_X\|_{\infty}$.}
\addtocounter{footnote}{1}
\footnotetext{We take $m=100k$. Hence, existing LLMs with much larger vocabulary size would produce negligible approximation error.}

\begin{thm}[Approximation of Max-min Detection Rate]
\label{thm:approx_maxmin_detection}
    Given $|S|=k$, $|\calX| = m$, and the inf-norm constraint $\lambda \in [0,1]$. Let $t = \floor{\frac{1}{\lambda}}$, and $Y\sim Bin(t,\frac{1}{k})$ An approximation of the optimal minimax detection probability is given by: 
    \begin{align}
    \tilde{R}^\star_d(\lambda) = 1 - \frac{1}{2k}- 
    \frac{1}{4}\left[\sum_{c=0}^t \Pr[Y=c] \left(\left|(c-t)\lambda + (1-\frac{1}{k})\right| + (k-1)\left|c\lambda -\frac{1}{k}\right|\right)\right]
    \end{align}
    The approximation error decays as $O(\frac{1}{m})$. Specifically:
    \begin{align}
        \left|\tilde{R}^\star_d(\lambda) - R^\star_d(\lambda)\right| \leq \frac{2k\ceil{\frac{1}{\lambda}}}{m}
    \end{align}
\end{thm}

We plot the results of Theorem \ref{thm:optimal_maxmin_detection} and \ref{thm:approx_maxmin_detection} in Figure \ref{fig:Rd_star}. For all $\lambda$ and $k$ values, the approximated maxmin detection coincides with the closed-form $R^\star_d(\lambda)$. We choose $m=100*k$. The overlap between the actual and approximated $R^\star_d(\lambda)$ in the plot testifies our result that the approximation error decays with $m$. In practice, since LLMs have a much large vocabulary, where $m\approx 100,000$\cite{grattafiori2024llama}, the approximation error will be negligible.

\section{Sequential Watermarking}
\label{sec:sequential}
While this paper focused on a single-shot analysis of token distribution watermarking, general text generation involves sequential prediction of long token sequences.
A common approach involves applying a token-level watermarking of the next token distribution and designing token-level test statistics.
This approach was shown to benefit from favorable performance \cite{he2024universally,kirchenbauer2023watermark,aaronson2023watermark}, albeit being theoretically suboptimal.
We note that our one-shot method readily extends to a sequential token-level scheme as we can treat each step as a one-shot problem, and considering an average test $\frac{1}{n}\sum \mathbf{1}[f_i(A_i,B^m_i) = S_i]$ which we them compare with some threshold $\tau\in[0,1]$.
We leave the theoretical analysis of the token-level extension of our scheme to future work, while showing empirical results in Section \ref{sec:exp}.
In the simplified case when $X^n$ are i.i.d., we provide the following bounds on the detection probability (a related result was given in \cite{chao2024watermarking} bounding mismatch proportion using entropy):
\begin{prop}
\label{prop:sequential_Rd_bound}
    Let $Q^n=Q^{\otimes n}_X$ be the an i.i.d. token distribution, let $S^n\sim P^{\otimes n}_S$ and apply the one-shot CC on each step $i\in[1:n]$, then
    \begin{equation*}
        1-2^{-\left(\frac{n}{2}+1\right)}\left(g(\tilde{p})\right)^n \leq \rd \leq \frac{1}{2}\left(1 + \sqrt{1-\left(\frac{\left(g(\tilde{p})\right)^2}{2}\right)^n}\right),
    \end{equation*}
    where $\tilde{p}=\min(\tilde{p}_0,\tilde{p}_1)$ is similarly defined as in the on-shot case, and $g(p) \triangleq p + \sqrt{\frac{1-p}{2}}\left(1+\sqrt{1-2p}\right), p\in[0,0.5].$
\end{prop}
The proof utilizes bounds on TV in terms of the Hellinger distance, which benefits from a tensortization. 

\section{Experimental Results}\label{sec:exp}
We numerically evaluate the CC watermark on a uniform source over $m=10$ tokens and binary side information.
We compare our results with the solution of an exact GUROBI-based numerical solution \cite{gurobi} of \eqref{eq:curve_opt_iid} and a popular watermarking scheme termed red/green list \cite{kirchenbauer2023watermark}.
Additional information on the numerical solution of \eqref{eq:curve_opt_iid} is given in Appendix \ref{appdx:additional_gurubi}.
The red/green watermark tilts $Q_X$ according to the value of some $\delta\in\RR_{\geq 0}$ which implicitly controls the downstream $\rp$.

\subsection{One-Shot Performance Analysis}
\begin{figure}[!t]
  \centering
  \begin{subfigure}[b]{0.47\textwidth}
    \centering
    \includegraphics[width=\textwidth]{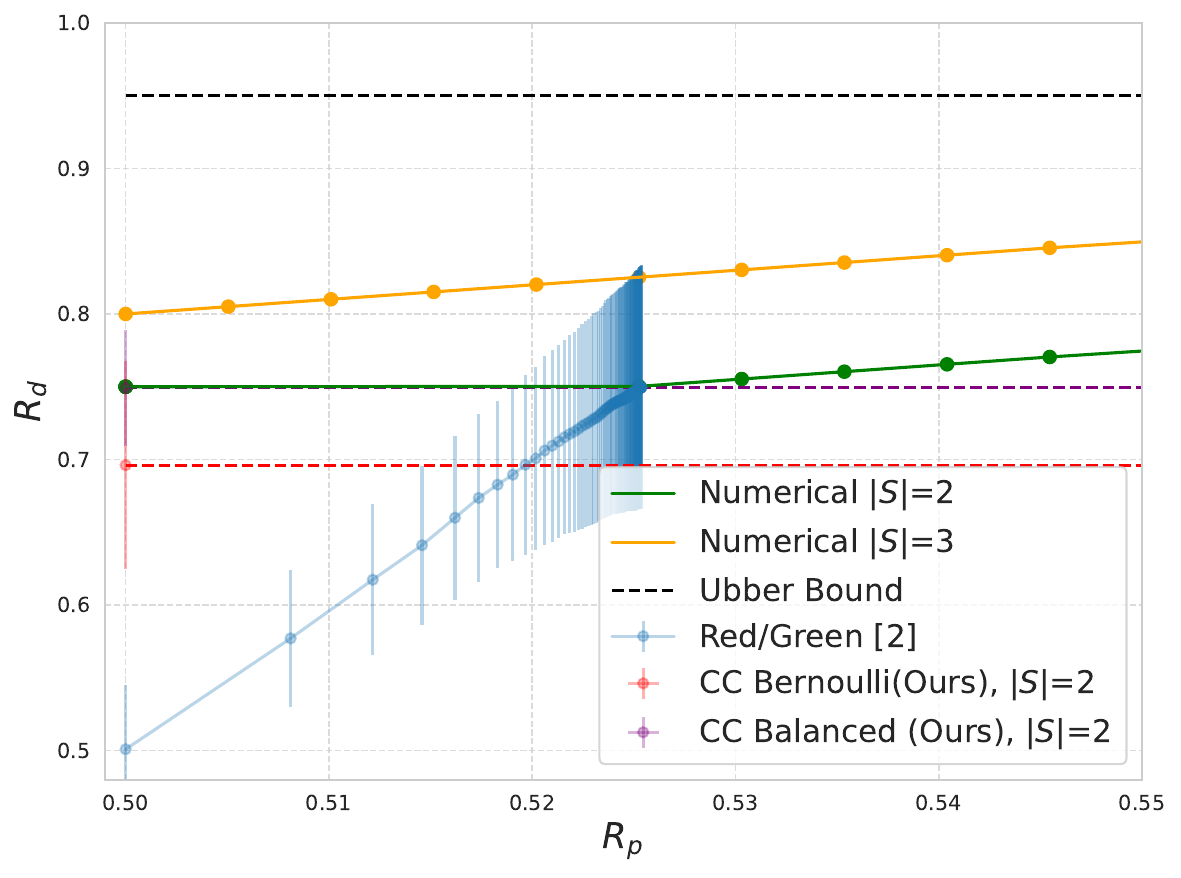}
    \caption{$Q_X=\mathsf{Unif}[1:m]$}
    \label{fig:exp_results_unif}
  \end{subfigure}
  \hspace{0.5cm}
  \begin{subfigure}[b]{0.47\textwidth}
    \centering
    \includegraphics[width=\textwidth]{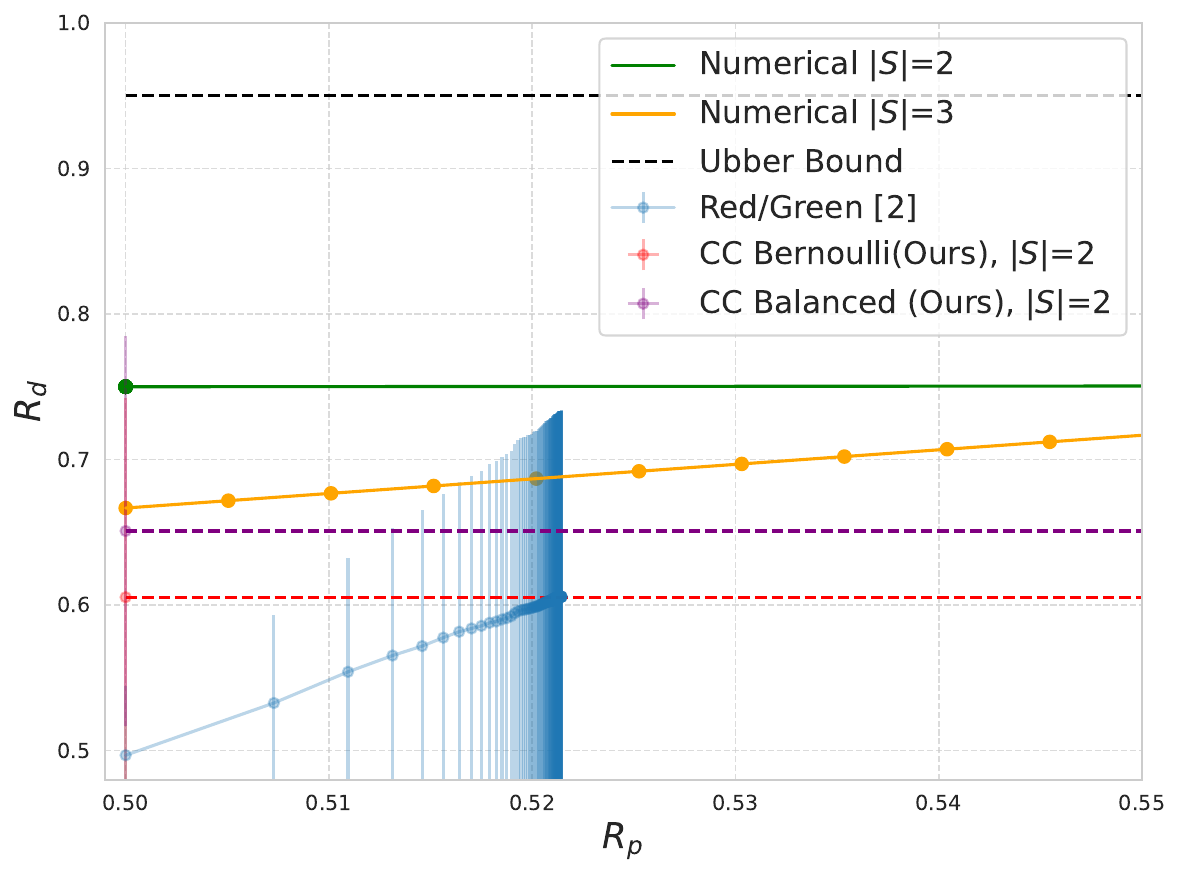}
    \caption{$Q_X=[0.5,0.5,0,\dots]$}
    \label{fig:subfig2}
  \end{subfigure}
  \caption{One-shot watermark detection results on $Q_X=\mathsf{Unif}(\cX)$. For $\alpha_p=0$, CC achieves a detection probability of $0.75$ and $0.7$ with balanced and Bernoulli partitions, respectively. CC Balanced achieves the optimal detection (Eq. \ref{eq:curve_opt_iid} with $\gamma = 1$ and $|\cS|=2$). Standard deviations plotted as two-sided bars.}
  \label{fig:rd_rp}
\end{figure}

\textbf{Detection-Perception Tradeoff: }
We present the $(\rd,\rp)$ tradeoff region for the one-shot watermarking setting. 
We consider the worst-case distribution within $\{Q_x, \|Q_X\|_\infty\leq\lambda\}$.
When $\lambda=\frac{1}{m}$, the resulting distribution is simply the uniform distribution over $\cX$ and when $\lambda\geq\frac{1}{2}$ it is given by a distribution with two nonzero entries valued $(\lambda, 1-\lambda)$. Such a distribution is a representative of a next-token distribution in the low entropy regime (highly predictable next token).
As seen from Figure \ref{fig:exp_results_unif}, for uniform $Q_X$, when we apply the CC scheme with $P_{B^m}$ sampled over balanced partitions, we obtain a gain of $\approx 0.07$ over sampling $B^m\stackrel{i.i.d.}{\sim}\mathsf{Ber}(\frac{1}{2})$, meeting the upper bound from \eqref{eq:curve_opt_iid}.
In contrast, the red-green detection coincides with ours in the limit of $\delta\to\infty$, intersecting with the suboptimal i.i.d. Bernoulli sampling method at $\delta\approx7.6$.
When $\delta=\frac{1}{2}$ we observe a decrease in the gain of sampling from the balanced partition sets.

\begin{figure}[!b]
  \centering
  \begin{subfigure}[b]{0.4\textwidth}
    \centering
    \includegraphics[width=\textwidth]{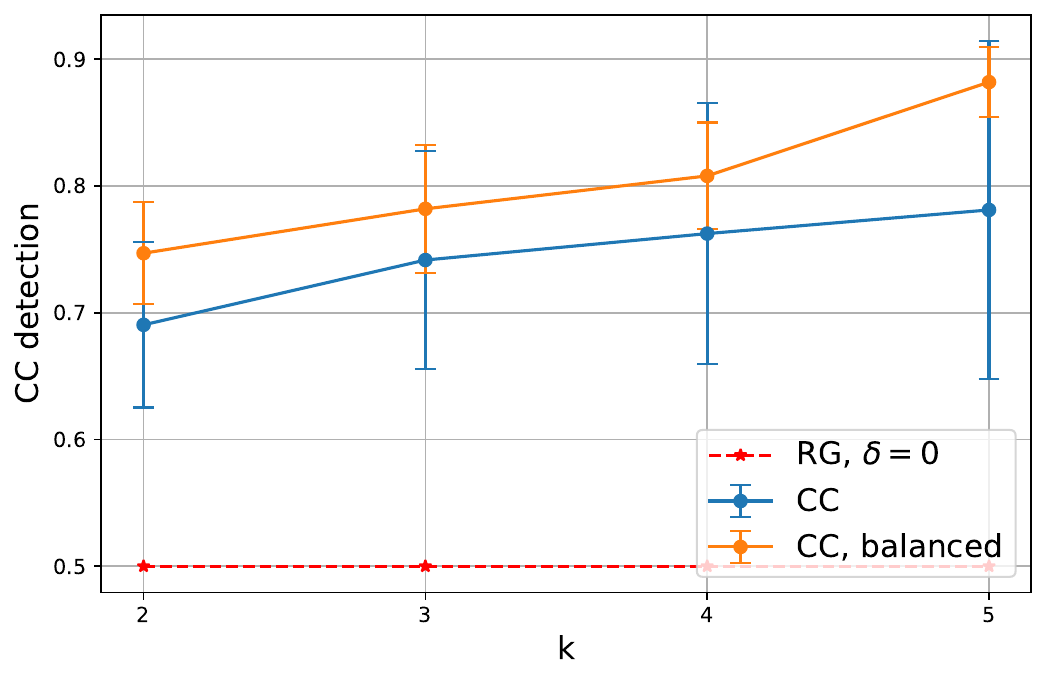}
    \caption{$m=10$}
    \label{fig:subfig11}
  \end{subfigure}
  \hspace{1cm}
  \begin{subfigure}[b]{0.4\textwidth}
    \centering
    \includegraphics[width=\textwidth]{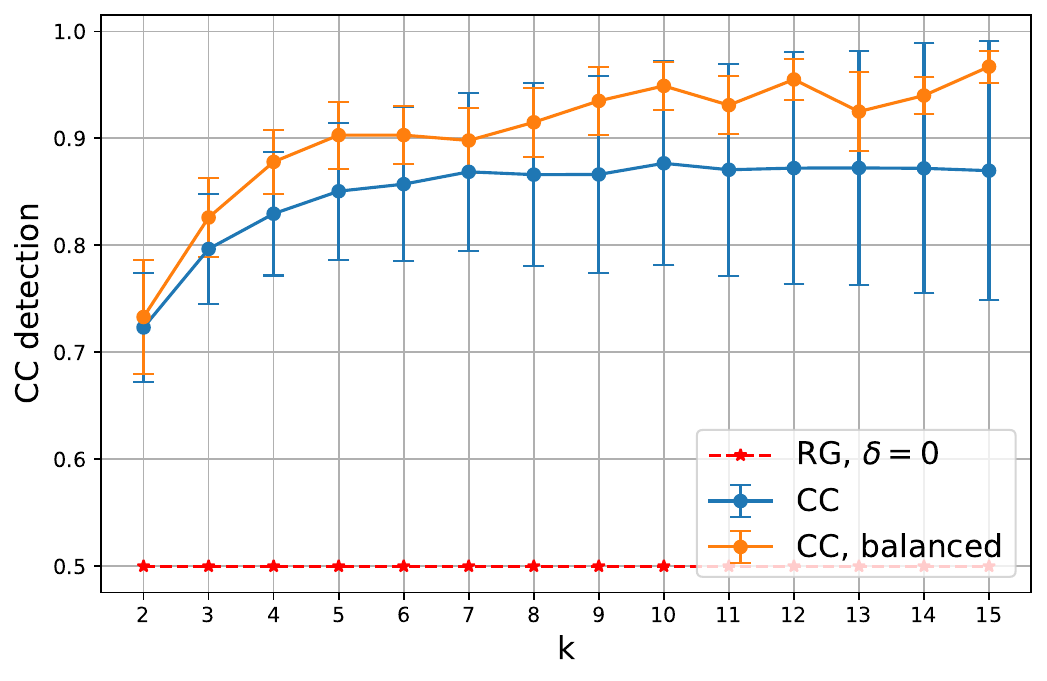}
    \caption{$m=60$}
    \label{fig:subfig22}
  \end{subfigure}
  \caption{Detection probability vs. $k$ for two values of $m$ and a uniform token distribution $Q_X$.}
  \label{fig:rd_vs_k}
\end{figure}

\textbf{Effect of $k$:}
Next, we analyze the effect of the side information alphabet size on the CC scheme performance.
We present a plot for $m=10$ which serves as an extension of the performance we present in Figure \ref{fig:exp_results_unif} and a plot for $m=60$, which allows us to further understand the effect.
As seen in Figure \ref{fig:rd_vs_k}, as $k$ increases, the detection rate of the CC watermark increases. However, the gain from increasing $k$ decreases as $k$ grows (or alternatively, as the ratio $m/k$ decreases). 
Furthermore, we note that the performance depends on the divisibility of $m$ by $k$; when $m/k$ is not an integer, we experience a degradation of performance. This follows from the inability to construct equally sized partitions of $\cX$, which, in turn, decreases the probability to result with a balanced partition.

\subsection{Sequential Watermarking}
\begin{figure}[!t]
  \centering
  \begin{subfigure}[b]{0.35\textwidth}
    \centering
    \includegraphics[width=\textwidth]{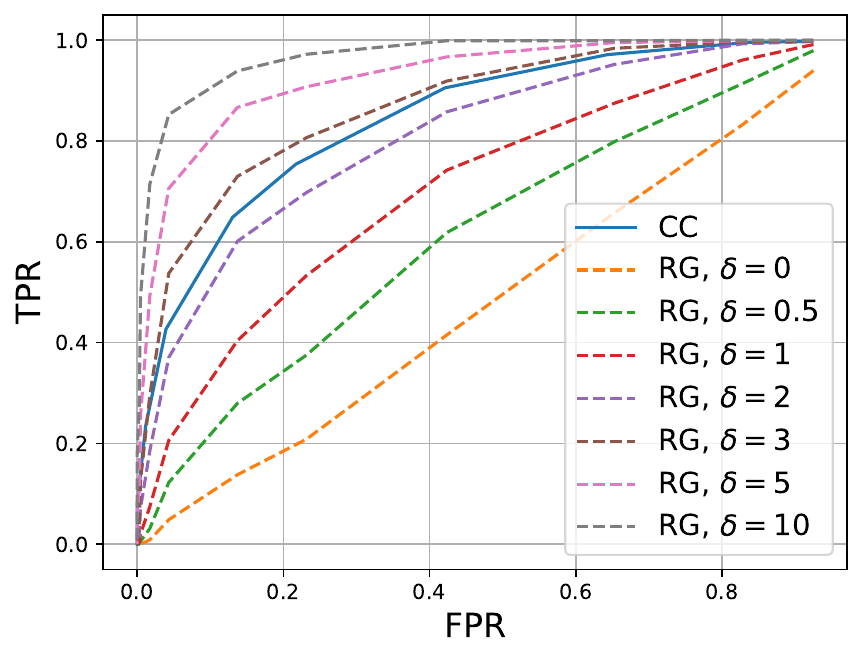}
    \caption{$\lambda=0.8$}
    \label{fig:subfig1_}
  \end{subfigure}
  \hspace{1cm}
  \begin{subfigure}[b]{0.35\textwidth}
    \centering
    \includegraphics[width=\textwidth]{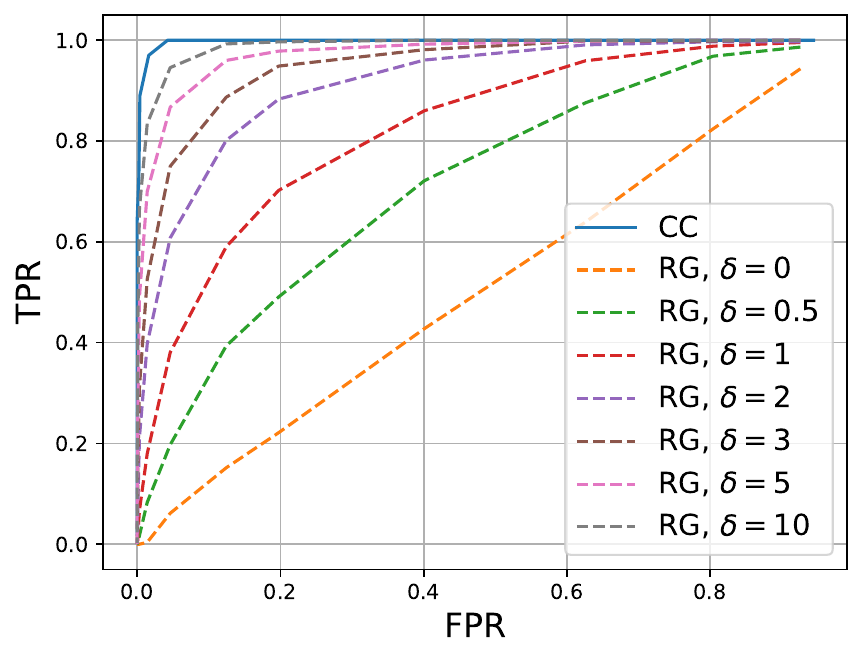}
    \caption{$\lambda=0.5$}
    \label{fig:subfig2_}
  \end{subfigure}
  \caption{ROC of the sequence-level watermarking scheme. We compare the red-green method \cite{kirchenbauer2023watermark} with the CC scheme (Section \ref{sec:cc_scheme}). We consider a range of $\delta$. An increase of $\delta$ increases detection, at the expense of higher perception (lower textual quality), while the CC method has fixed zero perception.}
  \label{fig:roc_deltas}
\end{figure}

We now present the performance of the CC watermark on a sequence level scheme. We present preliminary results on synthetically generated data, with the purpose of demonstrating the applicability of our method to a sequence-level test.
To that end, we consider the generation of $n$ tokens $A^n$, which are generated from a sequence of tokens $X^n\stackrel{i.i.d.}{\sim}Q_X$ using from $n$ i.i.d. samples of side information $s^n$ and randomness $(B^{m}(i))_{i=1}^n$.
We apply the token-level watermarking scheme to each element $X_i$ to generate $A_i$ and apply the following sequence-level threshold test
$$
r(A^n,S^n) = \left\{\frac{1}{n}\sum_{i=1}^n \mathbf{1}\left(f(A_i,B^m(i)) = S_i\right) \geq \tau\right\}
$$
for some threshold $\tau\in[0,1]$.
To understand the performance of the proposed sequence-level generalization, we analyze the ROC of the results scheme.
In out experiment, we consider $k=2$, $m=20$ and a sequence of $n=50$ tokens.
Figure \ref{fig:roc_deltas} compares the ROC of the CC scheme (sampling from balanced sets) with the red-green scheme for a range of $\delta$ values. 
We note that, while the CC method is perceptionless, it results in a better ROC than the red-green method.
Specifically, for $\lambda=0.5$, the CC method demonstrated better detection than the red-green method for the considered range of $\delta$ values. However, when $\lambda=0.8$, i.e., when the distribution is spikier, the red-green method with higher $\delta$ values result in a better ROC than the CC method, but at the cost of nonzero perception.


\begin{figure}[ht]
  \centering
  \begin{subfigure}[b]{0.35\textwidth}
    \centering
    \includegraphics[width=\textwidth]{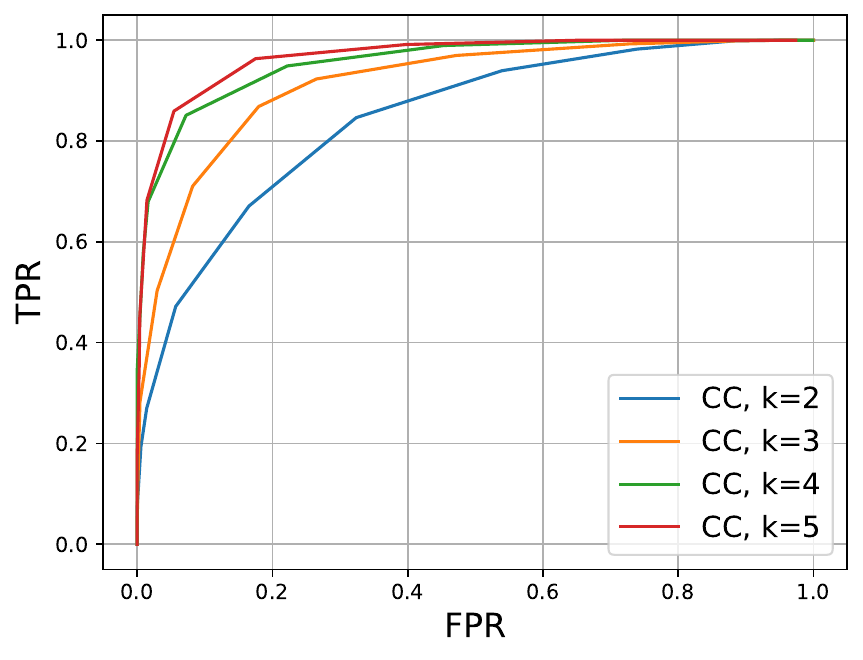}
    \caption{$Q_X$ has two nonzero values of $Q_1=0.8$ and $Q_2=0.2$.}
    \label{fig_}
  \end{subfigure}
  \hspace{1cm}
  \begin{subfigure}[b]{0.35\textwidth}
    \centering
    \includegraphics[width=\textwidth]{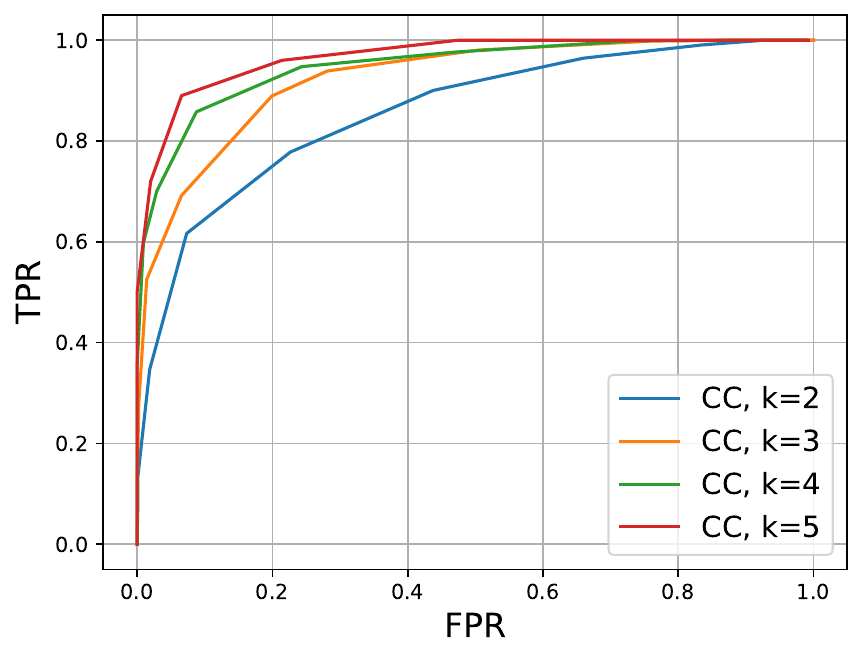}
    \caption{$Q_X$ has a spike of $\lambda=0.8$ and is uniform on the rest of its entries.}
    \label{fig:subfig22_}
  \end{subfigure}
  \caption{ROC of the sequence-level watermarking scheme under CC method for a range of $k$ values. }
  \label{fig:roc_vs_k}
\end{figure}

Finally, we analyze the effect of $k$ on performance in the sequential setting by observing the ROC for a range of $k$ values. 
Specifically, we consider $m=20$ and apply the sequential generalization of the CC watermark for $k\in\{2,3,4,5\}$.
We consider two distributions within the bounded infinity norm set with $\lambda=0.8$.
As can be seen in Figure \ref{fig:roc_vs_k}, as $k$ increases, the ROC improves. 
We note that, aligned with the results for the effect of $k$ in the one-shot setting, the significant improvement in the ROC occurs in the smaller $k$ regime, and higher $k$ values demonstrate diminishing returns.

\section{Watermarking Large Language Models}
In this section, we demonstrate the CC watermark on LLM text generation.
In this setting, the CC watermark is applied on a token-level, prior to each next token generation. That is, on each time-step $t$, the token distribution $Q_X$ is the model's next token prediction distribution given past tokens $Q(\cdot|x^{t-1})$.
Furthermore, the randomness $(S,B^m)$ is generated from the output of some hashing strategy that is similarly applied both on Alice and Bob's ends. The hashing function is often a function of a shared secret key and a subset of previously sampled tokens $(x_{t-1},\dots,x_{t-h})$ for some $0<h<t-1$.
To assess the detection performance of this watermarking method, we consider the $Z$-test, as considered in, e.g., \cite{kirchenbauer2023watermark,tu2023waterbench}. 
For a given statistic $Z\triangleq Z(x^T)$, the $Z$-test quantifies the deviation of Z from its mean and is usually compared against some threshold. 
Specifically, we reject the null hypothesis if we observe Z that is larger than a threshold at the tail of its distribution (e.g., at 95\% percentile).
As $Z$ corresponds to the sum of token-level Bernoulli random variables with probability $\gamma$, its sum is a binomial random variable, which, for a large enough sample, can be approximated with a Gaussian random variable by the Central Limit Theorem. 
Specifically, we have
    \begin{equation}
    \label{eq:zscore}
\bar{Z} \triangleq \frac{Z - T\gamma}{\sqrt{T\gamma(1-\gamma)}}\sim N(0,1).
    \end{equation}

For the RG watermark, we have the test statistic $Z_g=\sum_{i=1}^Tf(x_i,b^m)$ and $\gamma$ in this case represents the probability of green tokens. For the CC watermark we have the statistic $Z_{c}=\sum_{i=1}^T \mathbf{1}\{f(x_i,b^m)=s_i\}$, with $\gamma = \frac{1}{2}$ regardless of the probability of green tokens.
Note that under $H_0$, $Z_g$ and $Z_c$ have the same distribution for $\gamma = \frac{1}{2}$.
Hence, both are valid tests for the CC watermark and the two tests result in the same standard normal distribution under $H_0$ whenever the RG watermark partitions the vocabulary into balanced red and green lists ($\gamma = \frac{1}{2}$). 
The benefit of the $Z$-test is that it provides us with a $p$-value. 
As seen form \eqref{eq:zscore}, the larger the $Z$-score, the larger threshold we can allow, which translated into bigger true positive values at a given false positive constraint.
\begin{remark}
    Consider some $k\in \mathbb{N}$.
    For the CC watermark, Z corresponds to the sum of token-level Bernoulli random variables with probability $\frac{1}{k}$, assuming both  $f(x_i,b^m)$ and $s_i$ follows a categorical distribution uniform over $[k]$. The distribution of the detection test under the null is thus $\mathbf{1}\{f(x_i,b^m)=s_i\}\sim \text{Ber}(\frac{1}{k})$. Hence, the z-score for a general $k$ is identical to Eq.\eqref{eq:zscore} with $\gamma=\frac{1}{k}$
\end{remark}

\begin{table}[!b]
{
\setlength{\extrarowheight}{0.6ex}
\begin{tabular}{l||lll||lll}
\hline
\textbf{}              & \multicolumn{1}{l|}{$\mathsf{PPL}$ $\downarrow$}     & \multicolumn{1}{l|}{GM $\uparrow$}    & $Z$-score  $\uparrow$      & \multicolumn{1}{l|}{$\mathsf{PPL}$ $\downarrow$ }     & \multicolumn{1}{l|}{GM $\uparrow$}    & \multicolumn{1}{l}{$Z$-score $\uparrow$}       \\ \hline
\textbf{}              & \multicolumn{3}{l||}{\textbf{Knowledge Understanding (GM=32.76)}}     & \multicolumn{3}{l}{\textbf{Knowledge Memorization (GM=5.84)}}                            \\ \hline
\textbf{CC k=2 (ours)} & \multicolumn{1}{l|}{0.1594}  & \multicolumn{1}{l|}{31}    & 0.6448   & \multicolumn{1}{l|}{0.172}   & \multicolumn{1}{l|}{5.21}  & \multicolumn{1}{l}{0.6593}  \\ \hline
\textbf{CC k=3 (ours)} & \multicolumn{1}{l|}{0.1047}  & \multicolumn{1}{l|}{33.2}  & 0.051    & \multicolumn{1}{l|}{0.153}   & \multicolumn{1}{l|}{5.85}  & \multicolumn{1}{l}{0.575}   \\ \hline
\textbf{RG, d=1}       & \multicolumn{1}{l|}{0.2098}  & \multicolumn{1}{l|}{30}    & 1.37974  & \multicolumn{1}{l|}{0.2098}  & \multicolumn{1}{l|}{5.6}   & \multicolumn{1}{l}{0.6448}  \\ \hline
\textbf{RG, d=3}       & \multicolumn{1}{l|}{0.14977} & \multicolumn{1}{l|}{26.58} & 1.50712  & \multicolumn{1}{l|}{0.232}   & \multicolumn{1}{l|}{4.72}  & \multicolumn{1}{l}{1.459}   \\ \hline
\textbf{}              & \multicolumn{3}{l||}{\textbf{Finance QA (GM=21.35)}}                  & \multicolumn{3}{l}{\textbf{Longform QA (GM=21.88)}}                                      \\ \hline
\textbf{CC k=2 (ours)} & \multicolumn{1}{l|}{0.1634}  & \multicolumn{1}{l|}{21.32} & 2.477    & \multicolumn{1}{l|}{0.2346}  & \multicolumn{1}{l|}{21.17} & \multicolumn{1}{l}{3.393}   \\ \hline
\textbf{CC k=3 (ours)} & \multicolumn{1}{l|}{0.1352}  & \multicolumn{1}{l|}{21.16} & 2.954552 & \multicolumn{1}{l|}{0.19381} & \multicolumn{1}{l|}{21.76} & \multicolumn{1}{l}{4.1967}  \\ \hline
\textbf{RG, d=1}       & \multicolumn{1}{l|}{0.2044}  & \multicolumn{1}{l|}{21.51} & 1.5843   & \multicolumn{1}{l|}{0.2883}  & \multicolumn{1}{l|}{21.32} & \multicolumn{1}{l}{2.5836}  \\ \hline
\textbf{RG, d=3}       & \multicolumn{1}{l|}{0.2169}  & \multicolumn{1}{l|}{21.1}  & 4.331    & \multicolumn{1}{l|}{0.3063}  & \multicolumn{1}{l|}{21.04} & \multicolumn{1}{l}{5.86793} \\ \hline
\textbf{}              & \multicolumn{3}{l||}{\textbf{Multi-news (GM=26.18)}}                  & \multicolumn{3}{l}{\textbf{LCC (GM=51.51)}}                                              \\ \hline
\textbf{CC k=2 (ours)} & \multicolumn{1}{l|}{0.1331}  & \multicolumn{1}{l|}{25.86} & 1.62541  & \multicolumn{1}{l|}{0.2476}  & \multicolumn{1}{l|}{46.11} & \multicolumn{1}{l}{1.7249}  \\ \hline
\textbf{CC k=3 (ours)} & \multicolumn{1}{l|}{0.1073}  & \multicolumn{1}{l|}{26.07} & 2.19613  & \multicolumn{1}{l|}{0.2466}  & \multicolumn{1}{l|}{46.44} & \multicolumn{1}{l}{0.7195}  \\ \hline
\textbf{RG, d=1}       & \multicolumn{1}{l|}{0.16153} & \multicolumn{1}{l|}{26.45} & 1.40362  & \multicolumn{1}{l|}{0.3378}  & \multicolumn{1}{l|}{47.03} & \multicolumn{1}{l}{1.09314} \\ \hline
\textbf{RG, d=3}       & \multicolumn{1}{l|}{0.17816} & \multicolumn{1}{l|}{25.38} & 3.374    & \multicolumn{1}{l|}{0.3985}  & \multicolumn{1}{l|}{40.68} & \multicolumn{1}{l}{2.551}   \\ \hline
\end{tabular}
}
\vspace{0.3cm}
\caption{LLM Watermarking. We compare the CC watermark performance with the red-green (RG) method \cite{kirchenbauer2023watermark} on the WaterBench dataset. We consider six datasets, demonstrating various text generation tasks. We compare C with two values of $k$ and RG with two values of $\delta$, the tilting parameter. On each dataset, we report the baseline GM value which measures textual quality under no watermark.
}
\label{table:waterbench}
\end{table}
  
In order to compare our CC watermark method with the Red-Green watermark \cite{kirchenbauer2023reliability}, we follow the WaterBench benchmark \cite{tu2023waterbench} which considers six datasets covering common text generation tasks. The datasets consider short and long text generation tasks, free text generation and True/False questions, and coding tasks.
As previously described, to quantify detection we calculate $Z$-scores. 
As we cannot calculate perception in LLM text generation, we measure two proxies of it.
The first is the perplexity under the watermark. That is, for a sequence of generated tokens $x^T$, we calculate $-\log\mathsf{PPL}(x^t)=\frac{1}{T}\sum_{t=1}^T\log(Q_{X_t|X^{t-1},S_t}(x_t|x^{t-1},s_t))$. We interpret lower perplexity as better textual quality and denote it by PPL.
The second metric we report is WaterBench's generation metric (GM). 
For each of the text generation tasks considered, WaterBench proposes a lexicographical metric, which is a proxy for text generation; see \cite{tu2023waterbench} for more details.

We report the results in Table \ref{table:waterbench}. For \emph{generation quality}, CC attains the lowest $\mathsf{PPL}$ in every task, while maintaining a competitive GM score. It is improving upon the red-green method by up to \mbox{$45\%$} in $\mathsf{PPL}$ and up to \mbox{$+6.6$} GM points (Knowledge Understanding).
For \emph{detectability}, the CC detector achieves competitive or superior $Z$-scores: with $k{=}3$ it matches or exceeds the red-green's performance with $d{=}1$ on all tasks (peaking at $Z=4.2$ on Longform QA) while avoiding red-green's sharp quality drop; with $k{=}2$ it still yields statistically significant $Z$ values ($\approx0.6$–$3.4$) at virtually no fidelity cost.
Overall, CC offers a better tradeoff: maintaining textual quality and proposing improving detection capabilities.

\section{Conclusion}

This work presents a rigorous analysis of text watermarking in a one-shot setting through the lens of hypothesis testing with side information. 
We analyze the fundamental trade-off between watermark detection power and distortion in generated textual quality.
A key insight of our approach is that effective watermark design hinges on generating a coupling between the side information shared with the watermark detector and a random partition of the LLM vocabulary.
We develop a perfect perception watermarking scheme -- the Correlated Channel Watermark (CC).
Our analysis identifies the optimal coupling and randomization strategy under the worst-case LLM next-token distribution that satisfies a min-entropy constraint. 
Under the proposed scheme, we derive a closed-form expression of the resulting detection rate, quantifying the cost in a max-min sense.
The CC scheme offers a framework that can potentially accommodate additional objectives of LLM watermarking, such as robustness against adversarial manipulations and embedding capacity.
Additionally, we envision future work implementing the scheme for sequential watermarking and extending it to the positive-perception regime, where minor adjustments to token probabilities are permitted in exchange for superior detection.


\clearpage 

\bibliographystyle{unsrt}
\bibliography{bibliography}

\onecolumn
\appendix

\section{Proofs of Theoretical Results}

In this appendix, we include comprehensive overview of related works, as well as detailed proofs of our theoretical results, which are presented in the main body of the paper.

\subsection{Related Works}
Given the extensive volume of work in LLM watermarking, we focus our discussion on works that inform and contrast with our main contribution: theoretical frameworks for analyzing the limits of LLM watermarking.

\emph{Classical Information-Theoretic Approaches.} Post-process watermarking, where watermarks are embedded after content generation, has been extensively studied through information-theoretic lenses \cite{chen2000design,moulin2003information,martinian2005authentication}, particularly through the Gelfand-Pinsker (GP) channel \cite{gel1980coding,villan2006text,willems2000informationtheoretical}, which treats the LLM token $X\sim Q_X$ as the channel state for constructing the watermarked token. The GP scheme constructs auxiliary random variables $U\sim P(U|X)$ and encodes the watermarked token as $A=f(U,X)$. These approaches differ from our approach in two key aspects: (1) they typically require long sequences for joint typicality to hold, which leads to schemes that are intractable in the online setting with a large token vocabulary, while we focus on optimizing the one-shot minimax setting motivated by auto-regressive generation; and (2) they generally assume perfect knowledge of the underlying distributions, whereas our scheme is designed to work with the assumption that the underlying distribution is unknown, only the sampled token and side information are available.

\emph{Modern LLM Watermarking.} Kirchenbauer et al. \cite{kirchenbauer2023watermark} introduced the first watermarking scheme for LLMs, which divides the vocabulary into green and red lists and slightly enhances the probability of green tokens in the next token prediction (NTP) distribution. This seminal work sparked numerous developments \cite{aaronson2023watermark,he2024universally, bahri2024watermark,dathathri2024scalable,yang2023watermarking,ren2024subtle,huunbiased2024,zhao2024permute,chao2024watermarking,qu2024provably,xie2024debiasing,liuadaptive}, with several approaches focusing on distortion-free methods that maintain the original NTP distribution unchanged, e.g., \cite{kuditipudi2023robust,huunbiased2024,zhao2024permute,christ2024undetectable}. Unlike these methods which primarily focus on implementation strategies, our work provides a theoretical framework that characterizes optimal detection-perception trade-offs. Most related to our approach, Chao et al. \cite{chao2024watermarking} propose a watermark using optimal correlated channels, though our work differs by providing a complete characterization through joint optimization of the randomization distribution in the one-shot setting.

\emph{Theoretical Analysis of LLM Watermarking.} Recent work has advanced our theoretical understanding of LLM watermarking limitations. Huang et al. \cite{huang2023towards} designed an optimal watermarking scheme for a specific detector, but their approach requires knowledge of the original NTP distributions of the watermarked LLM, making it model-dependent. Li et al. \cite{li2024statistical} proposed detection rules using pivotal statistics, though their Type II error control relies on asymptotic techniques from large deviation theory and focuses on large-sample statistics, whereas our analysis addresses the fundamental one-shot case including explicit characterization of corner point cases and the development of an optimal correlated channel scheme. Most recently, He et al. \cite{he2024universally} characterizes the universal Type II error while controlling the worst-case Type-I error by optimizing the watermarking scheme and detector. In contrast to these approaches, we analyze optimal mean detection by formulating a minimax framework while balancing Type I and Type II errors through the use of an $E_\gamma$-information objective. In the minimax formulation, we provide the optimal mean detection in closed form and characterize the optimal distribution of randomness under adversarial token distributions. 

The development of the field is tracked through comprehensive benchmarks \cite{piet2023mark,tu2023waterbench,pan2024markllm,qiu2024evaluating} and surveys \cite{zhao2024sok, liu2024survey}.

\subsection{Proof for Proposition \ref{prop:tv_prob}}
\begin{proof}
Let us start by fixing $(P_S, Q_X,  Q_{X|S})$ and the priors $(\pi_0, \pi_1)$. Eve's hypothesis testing problem can be formulated as distinguishing between $H_0: A \sim Q_X$ and $H_1: A \sim \bar{Q}_X$. By the Neyman-Pearson Lemma, the optimal test statistic is given by the likelihood ratio $L(a) = Q_X(a)/\bar{Q}_X(a)$. The optimal decision rule takes the form $\delta(a) = \mathbbm{1}\{L(a) > \eta\}$ for some threshold $\eta$. The probability of correct detection for Eve can be expressed as:
\begin{align*}
\Pr(\hat{H}_E = C) &= \frac{1}{2}\Pr(\delta(A) = 1|H_1) + \frac{1}{2}\Pr(\delta(A) = 0|H_0)
\end{align*}

For the optimal threshold $\eta = 1$, this probability becomes:
\begin{align*}
\Pr(\hat{H}_E = C) &= \frac{1}{2} + \frac{1}{2}\sum_{a \in \mathcal{X}}|\bar{Q}_X(a) - Q_X(a)| \\
&= \frac{1}{2} + \frac{1}{2}\text{TV}(\bar{Q}_X, Q_X)
\end{align*}

Now, we turn to Bob's detection probability. Bob's hypothesis testing problem differs from Eve's due to his access to the side information $S$. His testing problem can be formulated as distinguishing between $H_0: (A,S) \sim Q_{X|S} \times P_S$ and $H_1: (A,S) \sim  Q_{X|S} \times P_S$. 

By the Neyman-Pearson Lemma, the optimal test statistic in this case is $L(a,s) = Q_{X|S}(a|s)/ Q_{X|S}(a|s)$. Given priors $(\pi_0, \pi_1)$ and let $\gamma = \frac{\pi_1}{\pi_0}$, the conditional probability of correct detection given $S=s$ is:
\begin{align}
\Pr(\hat{H}_B = C|S=s) &= \pi_0 \Pr(\delta(A) = 0|H_0) + \pi_1 \Pr(\delta(A) = 1|H_1)\\
&=  \pi_0 Q_{X|S}[L(a,s)\geq \gamma] + \pi_1  Q_{X|S}[L(a,s) \leq \gamma]\\
&= \pi_1 + \pi_0 Q_{X|S}[L(a,s)\geq \gamma] - \pi_1  Q_{X|S}[L(a,s) \geq \gamma]\\
&= \pi_1 + \pi_0 \left[ Q_{X|S}[L(a,s)\geq \gamma] - \frac{\pi_1}{\pi_0} Q_{X|S}[L(a,s) \geq \gamma \right]\\
&= \pi_1 + \pi_0 E_{\gamma}(Q_{X|S}|| Q_{X|S}).
\end{align}
The last equality comes from the alternative formula for $E_\gamma$ where $E_{\gamma}(P||Q) = \max_{\calA} [P(\calA)-\gamma Q(\calA)]$, and supremum is attained with $A = \{a|L(a,s)\geq \gamma\}$. 
\end{proof}

\subsection{Proof of Theorem \ref{thm:opt_cornerpoints}}\label{proof:opt_cornerpoints}
By the assumption of a uniform prior, we are looking for bounds on the quantity 
$\frac{1}{2}(1+E_\gamma( Q_{X|S}\|Q_X|P_S))$, which boils down to bounding $E_\gamma( Q_{X|S}\|Q_X|P_S) = \EE_S\left[E_\gamma( Q_{X|S}\|Q_X)\right]$.
First, note that under a uniform prior, this quantity is lower bounded by the performance of a random guess, i.e., $\frac{1}{2}\leq \rd$. In what follows, we develop an upper for $E_\gamma( Q_{X|S}\|Q_X|P_S)$.
For simplicity, denote $|\cX|=d$ and $|\cS|=m$.
Let $Q_{X|S=s_i}=p_i$ such that $p_1,...,p_m \in \Delta_d$, where $\Delta_d$ denotes the $d$-dimensional simplex.
Assume that $S\sim\mathsf{Unif}[m]$. Following the zero perception assumption, we have $\bar{Q}_X=Q_X$, i.e., $\frac{1}{m}\sum_{i=1}^m p_i = Q_X$.
Consequently, our TV-optimization, when jointly optimized also over the marginal distribution $Q_X$ is of the form:
\begin{equation}
\label{eq:opt}
\max_{p_1,...,p_m\in \Delta_d} \frac{1}{m}\sum_{i=1}^m \left\|p_i - \frac{\gamma}{m}\sum_{i=1}^m p_i\right\|_+,
\end{equation}
where $\|x\|_+\triangleq \sum_{i}(x_i)_+$
for $d\geq m$.
We are maximizing a convex function over a polytope, so the optimal solution lies on the extreme points. 
Thus $p_i= e_j$ for some $j\leq d$, where $e_j$ is the indicator vector with $j$-th entry equal to one. The problem boils down to determining how many times each vector $e_j$ shows up.

Denote with $q$ the probability vector corresponding to the distribution $Q_X$. We note that $q$ can be rewritten as
\begin{equation}\label{eq:q_def_proof_opt}
q \triangleq \frac{1}{m}\sum_{i=1}^m p_i = \frac{1}{m}\sum_{j=1}^d n_j e_j, 
\end{equation}
where $\sum_j n_j = m$ and $n_j\in\NN$. Denote the $j$-th entry of $q$ by $q_j$. We have $\|e_j - q\|_+ = (1-q_j)_+  = 1-q_j.$
Therefore:
\begin{align*}
\frac{1}{m}\sum_{i=1}^m \left\|p_i - \gamma q\right\|_+&\stackrel{a}{=}\frac{1}{m}\sum_{j=1}^d n_j \left\|e_j - \gamma q\right\|_+\\
&= \frac{1}{m}\sum_{j=1}^d n_j (1-\gamma q_j)_+\\
&\stackrel{b}{=} \sum_{j=1}^d q_j (1-\gamma q_j)_+
\end{align*}
where (a) follows from from rewriting the sum in terms of $e_j$ and (b) follows from the relation $q_j = \frac{n_j}{m}$, as can be seen from \eqref{eq:q_def_proof_opt} and by the definition of the indicator.
Out optimization problem had therefore boiled down to maximizing on the quantity 
\begin{align}\label{eq:thm1_opt}
\sum_{j=1}^d q_j (1-\gamma q_j)_+~\mbox{such that } q_j = k/m, k\in \mathbb{Z}, \sum_{j=1}^d q_j=1.
\end{align}
To solve \eqref{eq:thm1_opt}, we will examine various settings of the value of $\gamma$.
\subsubsection{$\gamma \leq 1$}
First, note that when $\gamma=0$ the objective sums up to $1$ by the constraints. 
Otherwise, note that whenever $\gamma\leq 1$, we have $(1-\gamma q_j)_+=1-\gamma q_j$. Thus, we have
$$
\sum_{j=1}^d q_j (1-\gamma q_j)_+ = 1 - \gamma\sum_{j=1}^n q_j^2.
$$
Thus, maximization of the objective, boils down to the minimization of the sum of squares.
We note that as $q$ is a probability vectors, the sum of square minimizes under the uniform distribution, with the minimum being $\frac{1}{m}$.
Thus, we have the upper bound 
$$
\frac{1}{2}(1+E_\gamma( Q_{X|S}\|Q_X|P_S))\leq \frac{1}{2}\left(1 + 1 - \frac{\gamma}{m}\right) = 1 - \frac{\gamma}{2m}.
$$
\subsubsection{$\gamma>1$}
In this case, we are not guaranteed with the positivity of  $(1-\gamma q_j)$. We will look for a strategy to choose the values of $(q_j)_j$ such that the considered sum is maximized, while not passing the threshold that nullifies the terms $(1-\gamma q_j)$.
For each $j$, denote each summand as $f(q_j)$, whose value is
$$
f(q_j) = \begin{cases}
    q_j - \gamma q_j^2,\quad q_j \leq \frac{1}{\gamma}\\
    0,\quad \text{else}.
\end{cases}
$$
Consequently, as $q_j$ is constrained to the set $(\frac{k}{m})_{k=0}^m$, whenever $\gamma\geq m$, no positive value of $q_j$ will result in a positive value of $f(q_j)$. Thus, the resulting sum is $0$, which implies that $\rd = \frac{1}{2}$.
Thus we will focus on $\gamma\in(1,m)$.
In this case, there is at least one possible value for each $q_j$ that results in a nonnegative value of $f(q_j)$.
First, we note that the mapping $x \mapsto x-\gamma x^2$ is a concave function of $x$ for $\gamma>0$, whose maximum is obtained in $x^\star = \frac{1}{2\gamma}$.
Therefore, we would like to set $q_j=\frac{1}{2\gamma}$ as this will maximize a single summand.
However, in most cases $\frac{1}{2\gamma}\notin(\frac{k}{m})_{k=1}^m$.
To that end, we will set the closes possible value to $\frac{1}{2\gamma}$ within the allowed
set. Second, we we would like to set as many $q_j$'s to the value $\frac{1}{2\gamma}$ while following the constraint $\sum_{j=1}^d q_j=1$, we will choose the lower value.
To summarize, for each interval $\frac{k}{m}\leq\frac{1}{2\gamma}\leq \frac{k+1}{m}$, we will set $q_j = \frac{k}{m}$.
The maximal amount of such $q_j$ we can set while following the sum constraint is $\lfloor \frac{m}{k} \rfloor $.
Thus, we have the following
\begin{align*}
    E_\gamma( Q_{X|S}\|Q_X|P_S) &= \left\lfloor \frac{m}{k} \right\rfloor\left( \frac{k}{m} - \gamma \left(\frac{k}{m}\right)^2\right)\\
    &\leq  1 - \frac{\gamma k }{m}.
\end{align*}
The corresponding bound on $\rd$ is $1 - \frac{\gamma k }{2m}$.
The bound is achievable whenever $m$ is divisible by $k$ within the resulting interval.
Note that the interval $\frac{k}{m}\leq \frac{1}{2\gamma}\leq \frac{k+1}{m}$ corresponds to the interval $\frac{m}{2(k+1)}\leq \gamma\leq \frac{m}{2k}$.
However, we already know the resulting bounds for $\gamma\geq m$ and $\gamma\leq1$. Thus, the relevant values of $k$ that correspond to this case are $k\in[1:\frac{m}{2}]$.
Finally, when $\frac{1}{2m}<\frac{1}{2\gamma}<\frac{1}{m}$ we cannot take the lower value $(k=0)$, and will therefore take higher value $k=1$.
However, note that $\frac{1}{2m}<\frac{1}{2\gamma}$ corresponds to $\gamma>m$. Thus, this sub-case $(\frac{1}{2m}<\frac{1}{2\gamma}\leq \frac{1}{m})$ boils down to $\gamma<\frac{m}{2}$ with corresponding upper bound of $1-\frac{\gamma}{m}$, which will merge with the interval $\gamma\leq 1$.
This concludes the proof $\hfill\square$

\subsection{Proof of Theorem \ref{thm:universal_ub}}
Let $Q_i\triangleq Q_{X|S=s_i}$ The proof follows from analyzing the following steps:
\begin{align*}
    \sup_{ Q_{X|S}}\sum_{s\in\cS} P_S(s) E_{\gamma}(\tilde{Q}_{X|S=s},Q_X) &= \sup_{ Q_{X|S}}\frac{1}{2|\cS|}\sum_{i=1}^{|\cS|}\|Q_i - \gamma Q_x \|_1\\
    &= \frac{1}{2|\cS|}\sup_{f:\cS\to\cX}\sum_{i=1}^{|\cS|}\|Q_{f(i)} - \gamma Q_x \|_1\\
    &\leq \frac{1}{2} \sup_{i\in \cX} \|Q_{i} - \gamma Q_x \|_1\\
    &=\sup_{i\in \cX} \left|1 - \gamma Q_x(i)\right|\\
    &= 1 - \gamma Q_{\mathsf{min}}
\end{align*}
Therefore, 
$$
\rd\leq \frac{1}{2}\left(1 + 1 - \gamma Q_{\mathsf{min}}\right) = 1 - \frac{\gamma Q_{\mathsf{min}}}{2}
$$
For the second equality, note that argmax of a convex function lies in the corner of the probability simplex.
$\hfill\square$

\subsection{Proof of Correlated Channel (CC) with Perfect Perception}
We prove that CC is a perfect perception scheme, i.e. $\EE_{S}\left[ Q_{X|S}\right](x) = Q_X(x)$.
Recall that $S=(Y,B^m)$.We have the following
\begin{align*}
    \EE_{S}\left[ Q_{X|S}\right](x) &= \sum_{y,b^m} \mu_{B^m}(b^m)P_Y(y)Q_X(x)\frac{P_{Y| Y}(y|f(x,b^m))}{P_Y(y)}\\
    &= Q_X(x) \sum_{y,b^m}\mu_{B^m}(b^m)P_{Y| Y}(y|f(x,b^m)).
\end{align*}
Denote by $\cB_1(x)\triangleq\left\{b^m: f(x,b^m)=1\right\}$ and denote $\cB_0(x)$ by the same token. We have
\begin{align*}
    &\EE_{S}\left[ Q_{X|S}\right](x) \\
    &=Q_X(x) \left(\sum_{b^m\in\cB_1(x)}\mu_{B^m}(b^m)\underbrace{\sum_{y=0,1}(b^m)P_{Y| Y}(y|1)}_{=1} + \sum_{b^m\in\cB_0(x)}\mu_{B^m}\underbrace{\sum_{y=0,1}\mu_{B^m}(b^m)P_{Y| Y}(y|0)}_{=1}\right)\\
    =Q_X(x).
\end{align*}
This concludes the proof.$\hfill\square$

\subsection{Proof of Proposition \ref{prop:optimal_coupling}}

By the dual representation of the total variation
	\begin{equation}
	\TV(P,Q) = \min_{P_{XY}}\{ \mathbb{P}[X \not = Y]: P_X = P, P_Y = Q\},
	\label{eq:tv-dual}
\end{equation}

Given $S\sim \mathsf{Unif}[k]$ and $P_{ Y}=\{p_1,...,p_k\}\in \Delta_k$. We have 
$\TV(P_S,P_{ Y}) = 1- \sum_{i=1}^k \min(\frac{1}{k},p_i)$.

We propose a coupling and shows that it achieves $\TV(P_S,P_{ Y})$. 

To simplify notation, let the distribution of $S$ and $ Y$ be $P$ and $Q$. Let $t  = \TV(P,Q)$. Assume that $0<t<1$. Define three probability distributions 
	 $R = \frac{P \wedge Q}{1-t}$, $P'=\frac{P-P \wedge Q}{t}$ and $Q'=\frac{Q-P\wedge Q}{t}$.
	Construct $P_{XY}$ as follows: 
	\begin{enumerate}
	\item Generate $B \sim \text{Bernoulli}(t)$. 
	\item If $B=0$, draw $Z \sim R$ and set $S= Y=Z$. 
	\item If $B=1$, draw $S\sim P'$ and $ Y \sim Q'$ independently. 
\end{enumerate} 

To show that this is a valid coupling, we verify the marginal distribution is kept the same.
We have: 
\begin{eqnarray*}
P_S(a) & = & \mathbb{P}(B=0)R(a) + \mathbb{P}(B=1)P'(a) \\
& = & (1-t) \left( \frac{P \wedge Q}{1-t}\right) (a) + t \left( \frac{P - P\wedge Q}{t}\right) (a) \\
& = & P(a)
\end{eqnarray*} 
Similarly, 
\begin{eqnarray*}
P_{ Y}(a) & = & \mathbb{P}(B=0)R(a) + \mathbb{P}(B=1)Q'(a) \\
& = & (1-t) \left( \frac{P \wedge Q}{1-t}\right) (a) + t \left( \frac{Q - P\wedge Q}{t}\right) (a) \\
& = & Q(a)
\end{eqnarray*} 
Therefore $P_{S Y}$ is a valid coupling.

Lastly, we show that for the specific coupling, $\mathbf{P}( Y \neq S)=\TV(P_S,P_{ Y})$

\begin{align*}
    \mathbf{P}( Y \neq S)&= 1-\mathbf{P}( Y=S)\\
    &= 1- (1-t)\\
    &= t\\
    &=\TV(P_S,P_{ Y})
\end{align*}

Thus, we have constructed a coupling $P_{S Y}$ that minimizes $\mathbf{P}( Y \neq S)$, which means that it maximizes $\mathbf{P}( Y = S)$. $\hfill\square$

\subsection{Proof of Remark \ref{thm:detection_test_optimal}}
The hypothesis test is the following: $H_0: X \sim Q_X$ and $H_1: X \sim Q_{X|S,B^m}$, where $Q_{X|S,B^m}$ is the CC-watermark distribution shown in equation \eqref{eq:wm_dist_channel}, and side information $S\sim \mathsf{Ber}(0.5)$.  We show $H_0$ is rejected by the CC detection test $S=f(X,B^m)$ if and only if it is also rejected by the likelihood ratio test (LRT).

If $H_0$ is rejected by CC detection test, then $S=f(X,B^m)$. Then, consider the likelihood ratio:
\begin{align}
    \frac{Q_X(X)}{Q_{X|B^m,S}(X)}&= 
    \frac{Q(X)}{Q_X(X)\frac{1}{P_S(S)}P_{S| Y}(S|f(X,B^m)}\\
    &= \frac{2}{P_{S| Y}(S|f(X,B^m)}\\
    &<1,
\end{align}
The density of $Q_{X|B^m,S}(X)$ follows from the CC-watermark, side information $P_S(S)=0.5$. The last inequality come from the Z-S channel construction: $\Pr_{S| Y}(S|f(S,B^m)\geq \frac{1}{2}$, if and only if $S=f(X,B^m)$. Since the likelihood ratio is less than $1$, $H_0$ is rejected by the LRT.

If $H_0$ is rejected by the LRT with threshold 1, then we have 
$$ \frac{Q_X(X)}{Q_{X|B^m,S}(X)} <1.$$
Expanding the likelihood ratio as above, this implies: 
$P_{S| Y}(S|f(X,B^m)<\frac{1}{2}.$
By construction of the Z-S channel, $S=f(X,B^m)$. Hence, $H_0$ is rejected by CC detection test.

\subsection{Proof of Proposition \ref{thm:detection_close_form}}\label{proof:tvq_as_tvy}
We start by proving the following identity:
    $$
    \TV\left(Q_X,Q_{X|(S,B^m)}|P_{S,B^m}\right) = \TV\left(P_{S},P_{S| Y}|P_{ Y}\right)
    $$

Proof:
Recall that in the correlated channel watermark we have side information $S$ and partition bits $B^m$. By definition, we have
\begin{equation}\label{eq:tv_tower}
    \TV(Q_X,Q_{X|S,B^m}|P_{S,B^m}) = \sum_{b^m}\sum_{s=0,1}\mu(b^m)P_S(s)\TV(Q_X,Q_{X|b^m,s}).
\end{equation}
Next, we simplify the TV expression within the sum. For any $(b^m,s)$ we have
\begin{align*}
    \TV(Q_X,Q_{X|(b^m,s)}) &= \sum_{x}\left| Q_X(x) - Q_X(x)\frac{P_{S| Y}(s|f(x,b^m))}{P_S(s)} \right|\\
    &= 2\sum_{x} Q_X(x)\left| \frac{1}{2} - p_{S| Y}(s| y)\right|,
\end{align*}
where recall that $ Y=f(X,B^m)$, $p_{S| Y}(s| y)$ is the corresponding coupling channel parameter, and
$S\sim \ber(\frac{1}{2})$.
We define the pre-image of $f$ for a fixed $b^m$ as $f^{-1}(\cdot,b^m):\{0,1\}\to 2^\cX$, with $f^{-1}(0),f^{-1}(1)\subseteq\cX$.
Plugging the simplified TV expression back into \eqref{eq:tv_tower}, we have 
\begin{align*}
    &\TV(Q_X,Q_{X|(b^m,s)}) \\
    &= \sum_{b^m}\mu(b^m)\sum_{s=0,1}\sum_{x} Q_X(x)\left| \frac{1}{2} - p_{S| Y}(s| y)\right|\\
    &= \sum_{b^m}\mu(b^m)\sum_{s=0,1} \left( \sum_{x\in f^{-1}(0,b^m)} Q_X(x)\left| \frac{1}{2} - p_{S| Y}(s|0)\right| + \sum_{x\in f^{-1}(1,b^m)} Q_X(x)\left| \frac{1}{2} - p_{S| Y}(s|1)\right|  \right)\\
    &= \sum_{b^m}\mu(b^m) \left( P_{ Y}(0)\sum_{s=0,1}\left| \frac{1}{2} - p_{S| Y}(y|0)\right| + P_{ Y}(1) \sum_{s=0,1}\left| \frac{1}{2} - p_{S| Y}(s|1)\right| \right)\\
    &=\TV\left(P_S,P_{S| Y} | P_{ Y} \right),
\end{align*}
where the randomness of $ Y$ is determined by the pair $(Q_X,\mu)$. This concludes the proof. $\hfill\square$

With this, we proceed to showing CC's detection rate. By Theorem \ref{thm:detection_test_optimal}, CC's detection rate is equal to that of likelihood ratio test. By Proposition \ref{prop:tv_prob} and under equal priors on TPR and TNR, we have 
\begin{align}
    R_d &= \frac{1}{2}(1+\TV(Q_X,Q_{X|S,B^m}|P_{S,B^m}))\\
    &= \frac{1}{2}\left(1+\TV(P_{S},P_{S| Y}|P_{ Y})\right),
\end{align}
where the last equality is due to the identity above.

Next, we obtain a closed form for $\TV(P_{S},P_{S| Y}|P_{ Y})$. By definition, we have
$$
\TV\left(P_{S},P_{S| Y}|P_{ Y}\right) = \tilde{p}_0 \TV\left(P_{S},P_{S| Y=0}\right)+\tilde{p}_1 \TV\left(P_{S},P_{S| Y=1}\right).
$$
Following Proposition \ref{prop:optimal_coupling}, the nature of the TV terms depends on wether $\tilde{p}_1\leq\frac{1}{2}$ or $\tilde{p}_0\leq\frac{1}{2}$ .
For $\tilde{p}_0\leq\frac{1}{2}$, the optimal coupling is given by a $Z$-channel, whose parameter is $\frac{2\tilde{p}_1-1}{2\tilde{p}_1}$. The TV terms are therefore given by
$$
    \TV\left(P_{S},P_{S| Y=0}\right) = \frac{1}{2}\left|\frac{1}{2}-1\right| + \frac{1}{2}\left|\frac{1}{2}\right| = \frac{1}{2}
$$
\begin{align*}
    \TV\left(P_{S},P_{S| Y=1}\right) &= \frac{1}{2}\left(\left|\frac{1}{2}-\frac{2\tilde{p}_1-1}{2\tilde{p}_1}\right| + \left|\frac{1}{2}-\frac{1}{2\tilde{p}_1}\right|\right)\\
    &= \frac{1}{2}\left(\left|\frac{1-\tilde{p}_1}{2\tilde{p}_1}\right| + \left|\frac{\tilde{p}_1-1}{2\tilde{p}_1}\right|\right)\\
    &= \frac{\tilde{p}_0}{2\tilde{p}_1}.
\end{align*}
Thus, we have
$$
\TV\left(P_{S},P_{S| Y}|P_{ Y}\right) = \tilde{p}_0.
$$
By the symmetry of the optimal coupling, for $\tilde{p}_1\leq\frac{1}{2}$ we have 
$$
\TV\left(P_{S},P_{S| Y}|P_{ Y}\right) = \tilde{p}_1.
$$

Hence, CC's detection rate is given by 
$R_d = \frac{1}{2}(1+\min(\tilde{p_0},\tilde{p_1}))$
. $\hfill\square$

\subsection{Proof of Theorem \ref{thm:optimal_maxmin_detection}}

We begin by proving Lemma \ref{lem:permutation_invariance}.
\subsubsection{Proof of Lemma \ref{lem:permutation_invariance}}
Let $\calS = [k]$ and $\calX=[m].$ For a given $Q_X = \bq = (q_1,\dots,q_m) \in \Delta_m$ and an $m$-length sequence $\mathbf{b}=(b_1,\dots,b_m)\in \calS^m$, we define the function $f:\calX \times \calS^m\to \calS$ as
\begin{equation}
    f(i,\mathbf{b}) = b_i.
\end{equation}
A sequence $\mathbf{b}$ induces a probability distribution  $\hat P(\mathbf{q},\mathbf{b}) $ over $\calS$ denoted as (with a slight abuse of notation)
\begin{equation}
    \hat P(s,\mathbf{q},\mathbf{b}) = \sum_{i=1}^m q_i \mathbf{1}\left[b_i=s\right]~\forall s\in[k].
\end{equation}
For a fixed $\mathbf{b}$ and $\bq$ and assuming that Alice uses the optimal coupling, Bob's probability of detection is given by the quantity
\begin{align}
    R_d(\bq, \mathbf{b})  &\triangleq  1 - \frac{1}{2}\mathsf{TV}\left(Q_S \| \hat P(\mathbf{q},\mathbf{b}) 
 \right) -\frac{1}{2k} \sum_{s=1}^k \hat P(s,\mathbf{q},\mathbf{b})\\
 &=  1-\frac{1}{2k} - \frac{1}{4} g(\bq,\mathbf{b}),
\end{align}
where
\begin{align}
 g(\bq,\mathbf{b}) \triangleq \sum_{s=1}^k \left|\hat P(s,\mathbf{q},\mathbf{b}) - \frac{1}{k}\right| 
\end{align}
where $Q_S$ is the uniform distribution. Our goal is to find a distribution over $P_{B^m}^*$ that maximizes the worst-case value of $R_d$ given a set of constraints on $\bq$.  Specifically, we analyze:
\begin{align}
    R_d^*(\lambda) &\triangleq \max_{P_{B^m}} \min_{\substack{\bq\in \Delta_m\\ \|\bq\|_\infty \leq \lambda}} \mathbb{E}\left[R_d(\bq,B^m)\right]\\
    &= 1 - \frac{1}{2k}-\frac{1}{4} \min_{P_{B^m}} \max_{\substack{\bq\in \Delta_m\\ \|\bq\|_\infty \leq \lambda}} \sum_{\mathbf{b}\in \mathcal{S}^m} P_{B^m}(\mathbf{b}) g(\bq, \mathbf{b}).
    \label{eq:rdstar}
\end{align}
The function
\begin{equation}
  H(P_{B^m}) \triangleq     \max_{\substack{\bq\in \Delta_m\\ \|\bq\|_\infty \leq \lambda}} \mathbb{E}\left[g(\bq, B^m) \right]
\end{equation}
is convex in the distribution $P_{B^m},$ since it is the maximum of linear functions. Let  $P_{B^m}^*$ be a distribution that minimized $H$ and consider the permutation $\pi:\mathcal{S}^m\to \mathcal{S}^m$, define $\tilde P_\pi(\mathbf{b}) = P_{B^m}^*(\pi \circ \mathbf{b})$.  

Since $ \mathbb{E}_{P_{B^m}^*}\left[g(\bq, B^m) \right] =  \mathbb{E}_{\tilde P_\pi}\left[g(\pi \circ \bq, B^m) \right]$ for all $\bq$,  $H(\tilde P_\pi)=H(P_{B^m})$ from the symmetry of the maximum. Hence, from the equality in \eqref{eq:rdstar} $F(\tilde P_\pi)=F(P_{B^m})$ for $F(P_{B^m})\triangleq \min_{\substack{\bq\in \Delta_m\\ \|\bq\|_\infty \leq \lambda}} \mathbb{E}_{P_{B^m}}\left[\rd(Q_X,B^m)\right]$.
$\hfill\square$

Next, we proceed with the proof of Theorem \ref{thm:optimal_maxmin_detection}.

Let $C=m!$ be the number of permutations of an $m$-length sequence, we have
\begin{equation}
    F\left(\frac{1}{C}\sum_\pi\tilde P_\pi  \right) \leq F(P_{B^m}^*).
\end{equation}
Consequently, it is sufficient to restrict the minimization in $P_{B^m}$ to distributions that assign equal probability mass to sequences that are identical up to a permutation.

Denote by $\mathcal{P}_m$ the partition of $\mathcal{S}^m$ into sets of sequences that are equal up to a permutation, with $|\mathcal{P}_m|=K$. For simplicity, we denote $\mathcal{P}_m = \left(\calB_{1},\dots,\calB_K \right)$ and refer to $\calB_i$ as a \emph{permutation class} (alternatively, we could have named it orbits or type classes). Then
\begin{align}
    \min_{P_{B^m}} F(P_{B^m}) = \min_{\bw \in \Delta_{K} } \max_{\substack{\bq\in \Delta_m\\ \|\bq\|_\infty \leq \lambda}} \sum_{i=1}^K \frac{w_i}{|\calB_i|} \sum_{\bb\in \calB_i}g(\bq,\bb).
\end{align}
Observe that $g(\bq,\bb)$ is convex in $\bq$ (since it is the absolute value of a linear function in $\bq$), and thus the inner maximum is achieved at a vertex of the feasible set. The vertices of the polytope $\left\{\bq \in \Delta_m \mid \|\bq\|_\infty \leq \lambda \right\}$ are permutations of the vector
$$ \bq^*_\lambda = (\lambda,\dots,\lambda, 1-t\lambda, 0,\dots,0 ),$$
where $\bq^*$ has (i) exactly $t$ entries equal to $\lambda$ and $t$ is the largest integer such that $t\lambda \leq 1$ (assuming $\lambda \leq 1$), (ii) one entry equal to $1-t\lambda$, and (iii) the remaining entries equal to 0. 

Since the vertices are identical up to a permutation, and for any permutation $\pi$
\begin{equation}
 \sum_{\bb\in \calB_i}g(\bq,\bb) =  \sum_{\bb\in \calB_i}g(\pi \circ \bq,\bb),
\end{equation}
it is sufficient to select a vertex of the form $\bq^*_\lambda$. Thus, 
\begin{equation}
    \min_{P_{B^m}} F(P_{B^m}) = \min_{\bw \in \Delta_{K} } \sum_{i=1}^K \frac{w_i}{|\calB_i|} \sum_{\bb\in \calB_i}g(\bq^*_\lambda,\bb),
\end{equation}
and it sufficient to consider the optimal distribution $P_{B^m}^*$ as a distribution that selects a $\bb$ uniformly over a \emph{single} permutation class in $\calP_m$; namely the one that maximizes $\frac{1}{|\calB_i|} \sum_{\bb\in \calB_i}g(\bq^*_\lambda,\bb).$

Next, we aim to characterize $R_d^*(\lambda)$ for different values of $\lambda$. We denote by $P_{\calB}$ the distribution resulting from drawing a sequence at random from the permutation class $\calB\in \calP_m$.

Our goal is to compute
\begin{equation}
\label{eq:g_expectation}
    \mathbb{E}\left[g(\bq_\lambda^*,B^m )\right] = \sum_{s=1}^k \mathbb{E}\left[ \left|\hat P(s,\mathbf{q}_\lambda^*,B^m) - \frac{1}{k}\right| \right]
\end{equation}
Recall that the optimal choice of $P_{B^m}$ is to draw sequences uniformly from a single permutation class. Assuming w.l.o.g. that $\mathcal{S}=[k]$, fix a sequence $\bb\in \calS^m$ with $n_i$ entries equal to $i$, $i\in [k]$. For example, if $k=2$, $n_1$ is the number of entries equal to 1 and $n_2$ is the number of entries equal to 2. Naturally, $\sum_{i=1}^k n_i=m$.

Now, for a fixed $s\in \calS$, we can write
\begin{equation}
    P(s,\mathbf{q}_\lambda^*,B^m) = \lambda \sum_{i=1}^t X_i + (1-t\lambda)X_{t+1},
\end{equation}
where $t = \lfloor 1/\lambda \rfloor$ and $X_i \defined \ones\left(B_i = s \right)$. We can expand the expectation in the lhs of \eqref{eq:g_expectation} as
\begin{align}
    \mathbb{E}\left[ \left|\hat P(s,\mathbf{q}_\lambda^*,B^m) - \frac{1}{k}\right| \right] &= \mathbb{E}\left[\mathbb{E}\left[ \left|\hat P(s,\mathbf{q}_\lambda^*,B^m) - \frac{1}{k}\right| \left| \sum_{i=1}^t X_i \right. \right] \right]\\
    & = \sum_{c=0}^t \Pr\left(\sum_{i=1}^t X_i = c \right)\left( \Pr\left( X_{t+1} = 1 \left| \sum_{i=1}^t X_i = c\right. \right)\left| c\lambda +(1-\lambda t) -\frac{1}{k} \right| \right. \\
    &\qquad\qquad\qquad\qquad\qquad\qquad+ \left. \Pr\left( X_{t+1} = 0 \left| \sum_{i=1}^t X_i = c\right. \right)\left| c\lambda -\frac{1}{k} \right| \right). \label{eq:full_form}
\end{align}
For our sampling without replacement strategy, we have
\begin{align*}
    \Pr\left(\sum_{i=1}^t X_i = c \right) &= \frac{\binom{n_s}{c}\binom{m-n_s}{t-c}}{\binom{m}{t}},\\
     \Pr\left( X_{t+1} = 1 \left| \sum_{i=1}^t X_i = c\right. \right) &= \frac{n_s-c}{m-t}.
\end{align*}
Plugging these expressions in, we have:
\begin{equation}
     \mathbb{E}\left[g(\bq_\lambda^*,B^m )\right] = \sum_{s=1}^k \sum_{c=0}^t \frac{\binom{n_s}{c}\binom{m-n_s}{t-c}}{\binom{m}{t}}\left( \left(\frac{n_s-c}{m-t} \right)\left| c\lambda +(1-\lambda t) -\frac{1}{k} \right| + \left(1-\frac{n_s-c}{m-t} \right)\left| c\lambda -\frac{1}{k} \right|  \right)
\end{equation}
When we have an equal number of elements of each kind in the permutation class and $m$ is divisible by $k$, i.e., $n_1 = \dots=n_k = m/k$, the expression simplifies to:
\begin{equation}
     \mathbb{E}\left[g(\bq_\lambda^*,B^m )\right] = k \sum_{c=0}^t \frac{\binom{m/k}{c}\binom{m-m/k}{t-c}}{\binom{m}{t}}\left( \left(\frac{ (m/k)-c}{m-t} \right)\left| c\lambda +(1-\lambda t) -\frac{1}{k} \right| + \left(1-\frac{ (m/k)-c}{m-t} \right)\left| c\lambda -\frac{1}{k} \right|  \right)
\end{equation}

We can simplify this even further in the special case that $\lambda = 1/k$. In this case, $t=k$, and we don't have to consider the special case of $X_{t+1}$ -- $\bq_\lambda^*$ has $k$ entries equal to $\lambda$. In this case, denoting $Z_k = \sum_{i=1}^k X_i$ \eqref{eq:full_form}, simplifies to:
\begin{align}
    \mathbb{E}\left[ \left|\hat P(s,\mathbf{q}_\lambda^*,B^m) - \frac{1}{k}\right| \right] 
    & = \frac{1}{k} \sum_{c=0}^k \Pr\left(Z_k = c \right)\left| c -1 \right|\\
    &=\frac{1}{k} \left(\Pr\left(Z_k = 0 \right) + \sum_{c=1}^k \Pr\left(Z_k = c \right)\left( c-1 \right)\right)\\
    &= \frac{1}{k} \left(2\Pr\left(Z_k = 0 \right)-1+ \mathbb{E}[Z_k]\right)\\
    &= \frac{2}{k}\Pr\left(Z_k = 0 \right)\\
    &= \frac{2}{k} \times \frac{ \binom{ (k-1)m/k }{k}}{\binom{m}{k}} 
\end{align}
and, consequently, we arrive at the elegant expression
\begin{equation}
    \mathbb{E}\left[g(\bq_\lambda^*,B^m )\right] = 2\times \frac{ \binom{ (k-1)m/k }{k}}{\binom{m}{k}}.
\end{equation}

Hence, for any given $m,k,\lambda$, that satisfies $\lambda = \frac{1}{k}$ and $m$ divisible by $k$, we have (following Eq. \eqref{eq:rdstar}):
\begin{align}
    R_d^*(\lambda) 
    &= 1 - \frac{1}{2k}-\frac{1}{4} \mathbb{E}\left[g(\bq_\lambda^*,B^m )\right]\\
    &= 1 - \frac{1}{2k}-\frac{1}{2} \frac{ \binom{ (k-1)m/k }{k}}{\binom{m}{k}}
    \label{eq:rdstar_}
\end{align}

For $1/2\leq \lambda < 1$, $\bq^*_\lambda$ has two non-zero entries equal to $\lambda$ and $1-\lambda$. Consequently, $\hat P(\bq^*_\lambda,\bb)$ assigns probability 1 to one value of $S$ if $b_1=b_2$, otherwise assigns mass $1-\lambda$ and $\lambda$ to two separate values of $s$. Thus for a fixed distribution $P_\calB$ 
    \begin{equation}
    \label{eq:Rd_k2}
    \mathbb{E}_{P_\calB}\left[R_d(\bq^*_\lambda,B^m )\right] =1-\frac{1}{2k}- \Pr(B_1=B_2)\times \frac{k-1}{2k} -\frac{1}{4}  \Pr(B_1\neq B_2)\times \left(1-\frac{2}{k} + \left|\lambda-\frac{1}{k}\right| + \left|1-\lambda - \frac{1}{k}\right| \right).
    \end{equation}
    We need to select the set $\calB$ that maximizes $\Pr(B_1\neq B_2)$. For $m$ even and $k=2$ (i.e., $\calS$ binary), $\calB$ is the permutation class of the sequence of equal number of each element, we have
    $\Pr(B_1 = B_2) =   \frac{m-2 }{2(m-1)} $, $\Pr(B_1 \neq B_2) = \frac{m }{2(m-1)}$, which simplifies $R_d(\lambda)^*$ to 
    \begin{equation}
    \label{eq:binary-bound}
         R_d^*(\lambda) = \frac{3}{4}- \frac{m\lambda -1}{4(m-1)}~\mbox{for} ~ k=2,~\frac{1}{2}\leq \lambda \leq 1. 
    \end{equation}
As $m\to \infty$, $R^\star_d(\lambda) \to \frac{3}{4}-\frac{m}{4}$.

\begin{remark} 
We now clarify the structure of $\calB^*$ showing it contains exactly those binary elements with balanced occurrence of 0's and 1's, i.e., $\calB^* = \{b^m:\text{equal number of 0's and 1's}\}$, when $\frac{1}{2}\leq \lambda <1$, $k=2$ and $m$ is even. For $S = \{0,1\}$, i.e. $k=2$, permutation classes are characterized by the number of 1$'s$. Let $\alpha$ be the number of $1's$ in $\calB$ and $m-\alpha$ be the number of $0$'s. 
From Eq \eqref{eq:Rd_k2}, we need to select the set $\calB$ that maximizes $\Pr(B_1\neq B_2)$:
\begin{align}
    \alpha^* = \argmax_{\alpha\in [m]} \Pr[B_1\neq B_2] = \argmax_{\alpha \in [m]} 2\frac{\alpha(m-\alpha)}{m(m-1)} = \frac{m}{2}.
    \label{eq:alpha*_k2}
\end{align}
   
\end{remark}

Next, we consider the case for $\frac{1}{3}\leq\lambda < \frac{1}{2}$. $\bq^*_\lambda$ has three non-zero entries: $\bq^*_\lambda = (\lambda, \lambda, 1-2\lambda, 0, ...,0)$. Consequently, there are 4 cases with the corresponding $\hat{P}(\bq^*_\lambda,\bb)$ and $g(\bq^*_\lambda,\bb)$: 
\begin{align*}
    a. & B_1 = B_2 = B_3: \quad \hat{P}= [1, 0, ...,0] \quad g(\bq^*_\lambda,\bb) = 2(1-\frac{1}{k})\\
    b. & B_1 = B_2, B_3: \neq B_1 \quad \hat{P}= [2\lambda, 1-2\lambda, 0 ...,0]  \quad g(\bq^*_\lambda,\bb) = (2\lambda - \frac{1}{k})+|1-2\lambda -
    \frac{1}{k}| + \frac{1}{k}(k-2)\\
    c. & B_1 \neq B_2, B_3 = (B_1 \vee B_2): \quad \hat{P}= [1-\lambda, \lambda, 0 ...,0] \quad  g(\bq^*_\lambda,\bb) = |\lambda - \frac{1}{k}|+|1-\lambda -
    \frac{1}{k}| + \frac{1}{k}(k-2)\\
    d. & B_1 \neq B_2 \neq B_3: \quad \hat{P}= [\lambda, \lambda, 1-2\lambda, 0 ...,0] \quad g(\bq^*_\lambda,\bb) = 2|\lambda - \frac{1}{k}|+|1-2\lambda -
    \frac{1}{k}| + \frac{1}{k}(k-3)
\end{align*}
Recall that to maximize $\mathbb{E}_{P_\calB}\left[R_d(\bq^*_\lambda,B^m )\right]$, we need to minimize $\mathbb{E}_{P_\calB}\left[g(\bq^*_\lambda,B^m )\right]$. 

For k=2, case $d$ is invalid and case $c$ produces the minimum $g(\bq^*_\lambda,\bb)$. Hence, we select the set $\calB$ that maximizes $\Pr[B_1 \neq B_2, B_3 = (B_1 \vee B_2)]$, which is equivalent to maximizing $\Pr[B_1 \neq B_2]$. Following \eqref{eq:alpha*_k2}, $\calB^* = \{b^m:\text{equal number of 0's and 1's}\}$. We have $\Pr[B_1=B_2=B_3]= \frac{m-4}{4(m-1)}$, 
$\Pr[B_1=B_2, B_3\neq B_1]= \frac{m}{4(m-1)}$ and $\Pr[B_1\neq B_2, B_3 = (B_1 \vee B_2)]= \frac{m}{2(m-1)}$.

The resulting $R_d(\lambda)^*$ is:
\begin{align}
    R^*_d(\lambda) = \frac{3}{4} - \frac{m-2}{8(m-1)} ~\mbox{for} ~ k=2,~\frac{1}{3}\leq \lambda <\frac{1}{2}.
\end{align}
As $m\to \infty$, $R^*_d(\lambda)\to \frac{5}{8}$.


\subsection{Proof of Theorem \ref{thm:approx_maxmin_detection}}

Our results so far have been based on the discussion that it is sufficient to consider the optimal distribution $P_{B^m}^*$ as one that selects $\bb$ uniformly over a single permutation class $\calB^* \in \calP_m$. Recall that $\bb$ is a sequence of $m$ elements each take a value in $\calS$: $|\bb|=m$ and $\calS=k$. Recall as well that $\calB$ can be characterized by the proportion of each element of $S$: for $i \in [k]$, denote the proportions as $[p_1,...,p_k]$, where $$p_s = \frac{\sum_{i=1}^m \ones[\bb_i=s]}{m} \quad \forall \bb \in\calB.$$ 
Hence, sampling an $\bb$ uniformly over $\calB^*$ can be equivalently defined as the following process: given $m$ elements with predefined proportions $[p_1,...,p_k]$, sample $m$ times with replacement.

To generalize the analysis for other ranges of $\lambda$, $k$, and $m$, we consider an alternative process in which rather than fixing the proportions over $m$ elements, we take $[p_1,...,p_k]$ as probabilities. $\bb$ amounts to $m$ i.i.d samples from a categorical distribution: $\bb_i \stackrel{i.i.d}{\sim} \textsc{Categorical}(p_1,...,p_k)$. Recall that optimal $B^*$ amounts to having an equal number for each element in $\calS$. Hence,  for all $i\in [k]$, $p^*_i = \frac{1}{k}$. 

Furthermore, recall that the adversarial distribution for a given min-entropy constraint $\lambda$ is: $\bq^* = [\lambda, \lambda, ..., 1-t\lambda,0,...,0]$, where $t = \lfloor{\frac{1}{\lambda}}\rfloor$. For the purpose of characterizing $\mathbb{E}_{P_\calB} g(\bq^*,\bb)$, only the color of the first $t+1$ draws matter, because the rest have 0 probabilities. 

Let $X_i \defined \ones\left(B_i = s \right)$, for a fixed $s\in \calS$. $X_i \stackrel{i.i.d}{\sim}\textsc{Ber}(\frac{1}{k})$. We can compute $\mathbb{E}_{P_\calB} g(\bq^*,\bb)$ in closed form. Following \eqref{eq:g_expectation} and \eqref{eq:full_form}, for sampling with replacement, we have:  

\begin{align}
\mathbb{E}\left[g(\bq_\lambda^*,B^m )\right] 
    &= \sum_{s=1}^k \mathbb{E}\left[ \left|\hat P(s,\mathbf{q}_\lambda^*,B^m) - \frac{1}{k}\right| \right]\\
   &= \sum_{s=1}^k \sum_{c=0}^t \Pr\left(\sum_{i=1}^t X_i = c \right)\left( \Pr\left( X_{t+1} = 1 \left| \sum_{i=1}^t X_i = c\right. \right)\left| c\lambda +(1-\lambda t) -\frac{1}{k} \right| \right. \\
    &\qquad\qquad\qquad\qquad\qquad\qquad+ \left. \Pr\left( X_{t+1} = 0 \left| \sum_{i=1}^t X_i = c\right. \right)\left| c\lambda -\frac{1}{k} \right| \right)\\
    &= k \sum_{c=0}^t \Pr[Y=c] \left(\frac{1}{k} \left|c\lambda + (1-\lambda t)-\frac{1}{k}\right| + (1-\frac{1}{k})\left|c\lambda -\frac{1}{k}\right|\right)\\
    &= \sum_{c=0}^t \Pr[Y=c] \left(\left|(c-t)\lambda + (1-\frac{1}{k})\right| + (k-1)\left|c\lambda -\frac{1}{k}\right|\right)
\end{align}
where $Y\sim Bin(t,\frac{1}{k})$, and hence $\Pr[Y=c] = \binom{t}{c} (\frac{1}{k})^c (1-\frac{1}{k})^{t-c}$

By Eq. \ref{eq:rdstar}, the approximated minimax detection is given by: 
    \begin{align}
    \tilde{R}^\star_d(\lambda) = 1 - \frac{1}{2k}- 
    \frac{1}{4}\left[\sum_{c=0}^t \Pr[Y=c] \left(\left|(c-t)\lambda + (1-\frac{1}{k})\right| + (k-1)\left|c\lambda -\frac{1}{k}\right|\right)\right]
    \end{align}



Finally, we analyze the approximation error of $\tilde{R}^\star_d(\lambda)$. Define $H_\bb$ and $M_\bb$ as the distribution of $\bb$ when we sample without (which yields $R^\star_d(\lambda) $) and with replacement (which yields $\tilde{R}^\star_d(\lambda)$). First, notice that $g(\bq^*,\bb) \leq \frac{2(k-1)}{k}\leq 2$ by considering the TV between singleton distribution and uniform. Then, by triangular inequality, we have: 

\begin{align}
    \left|\tilde{R}^\star_d(\lambda) - R^\star_d(\lambda)\right|
    &= \frac{1}{4} \left|(\mathbb{E}_{\bb \sim H_\bb} g(\bq^*,\bb) -\mathbb{E}_{\bb \sim M_\bb} g(\bq^*,\bb))\right| \\
    &= \frac{1}{4}\left|\sum_{\bb} g(\bq^*,\bb) (H_\bb(\bb)- M_\bb(\bb))\right|\\
    &\leq \frac{1}{4}*2 \left|\sum_{\bb}  (H_\bb(\bb)- M_\bb(\bb))\right|\\
    &\leq \frac{1}{2} \sum_{\bb} \left| (H_\bb(\bb)- M_\bb(\bb))\right|\\
    &= \TV(M_\bb, H_\bb)\\
    &\leq \frac{2k\ceil{\frac{1}{\lambda}}}{m}
\end{align}
The last inequality follows from de Finetti's Finite Exchangeable Sequences\cite{diaconis1980finite}.

\subsection{Proof of Proposition \ref{prop:sequential_Rd_bound}}
Let $n<\infty$ and assume that $X^n\sim Q^{\otimes n}$, $S^n\sim P^{\otimes n}$ and $(B^m_i)_{i=1}^n\sim P^{\otimes n}_{B^m}$.
Consequently, the CC watermarked distribution is also i.i.d. distributed according $\tilde{Q} =  Q_{X|S}$.
On Bob's end, the detection probability is given by the expression
$$
\rd = \frac{1}{2}\left(1+ \TV\left( (PQ)^{\otimes n}, (P\tilde{Q})^{\otimes n}\right) \right),
$$
where $P\tilde{Q}(S,X)=P(S)\tilde{Q}(X|S)$
To that end, we focus on obtaining bounds on the aforementioned TV term. For a pair of distributions $P,Q$, we have the following Hellinger bounds on the TV distance  \cite{polyanskiy2024information}:
\begin{equation}\label{eq:hellinger_bounds}
    \frac{1}{2}H^2(P,Q) \leq \TV(P,Q) \leq H(P,Q)\sqrt{1-\frac{1}{4}H^2\left(P,Q\right)},
\end{equation}
where, for two measures $P,Q$ on a finite alphabet $\cX$, the squarred Hellinger divergence is given by the following equivalent forms
$$
H^2(P,Q)\triangleq \EE_Q\left[\left(1 - \sqrt{\frac{P}{Q}}\right)^2\right]=\sum_{x\in\cX}\left(\sqrt{P(x)}-\sqrt{Q(x)}\right)^2=2-2\sum_{x\in\cX}\sqrt{P(x)Q(x)}.
$$
For a pair of product distributions $(P^{\otimes n},Q^{\otimes n})$, the squarred Hellinger divergence benefits from the relation \cite{polyanskiy2024information}
$$
H^2\left(P^{\otimes n},Q^{\otimes n}\right) = 2-\left(1-\frac{1}{2}H^2(P,Q)\right)^n.
$$
Our problem therefore boils down to characterize $H^2\left(PQ,P\tilde{Q}\right)$.
We have
\begin{align*}
    H^2\left(PQ,P\tilde{Q}\right) &= \sum_{x,s}P(s)\left(\sqrt{Q(x)}-\sqrt{\tilde{Q}(x|s)}\right)^2\\
    &= \EE_S\left[H^2(Q(X),Q(X|S))\right].
\end{align*}
For a given $s,b^m)$,we have
\begin{align*}
    H^2(Q(X),Q(X|S=s) &= 2 - 2\sum_{x}\sqrt{Q(x)\tilde{Q}(x|s)}\\
    &= 2 - 2\sum_{x}Q(x)\sqrt{\frac{P_{S| Y}(s| y(x,b^m))}{P(s)}}\\
    &= 2\EE_{X}\left[1 - \sqrt{\frac{P_{S| Y}(s| Y(X,b^m))}{P(s)}}\right],
\end{align*}
where $P(S| Y)$ is the correlated channel.
Assuming $S\sim\mathsf{Ber}\left(\frac{1}{2}\right)$, we have
\begin{align*}
    H^2\left(PQ,P\tilde{Q}\right) &= 2\EE_{S,X}\left[1-\sqrt{\frac{P_{S| Y}(S| Y(X,b^m))}{P(S)}}\right]\\
    &= \EE_{ Y}\left[ 1 -\sqrt{2P(0| Y)}\right] + \EE_{ Y}\left[1 -\sqrt{2P(1| Y)}\right]\\
    &= 2 - \sqrt{2}\EE_{ Y}\left[P(0| Y)+P(1| Y)\right]\\
    &= 2- \sqrt{2}\left(\tilde{p}_0\left(\sqrt{p(0|0)}+\sqrt{p(1|0)}\right) + \tilde{p}_1\left(\sqrt{p(0|1)}+\sqrt{p(1|1)}\right) \right),
\end{align*}
where $ Y\sim\mathsf{Ber}(\tilde{p}_0,\tilde{p}_1)$.
Due to the symmetry of the correlated channel, we have for $\tilde{p}\triangleq\min(\tilde{p}_0,\tilde{p}_1)$
$$
H^2\left(PQ,P\tilde{Q}\right) = 2-\sqrt{2}f(\tilde{p})
$$
where
$$
f(\tilde{p})\triangleq \tilde{p} + \sqrt{\frac{1-\tilde{p}}{2}}\left(1 + \sqrt{1-2\tilde{p}}\right),
$$
which implies that
$$
H^2\left(P^{\otimes n},Q^{\otimes n}\right) = 2 - 2^{1-\frac{n}{2}}\left(f(\tilde{p})\right)^n.
$$
The bounds on the detection probability then follow by plugging the squarred Hellinger distance into \eqref{eq:hellinger_bounds}.$\hfill\square$

\subsection{Proof of Proposition \ref{prop:sequential_Rd_bound}}
Let $n<\infty$ and assume that $X^n\sim Q^{\otimes n}$, $S^n\sim P^{\otimes n}$ and $(B^m_i)_{i=1}^n\sim P^{\otimes n}_{B^m}$.
Consequently, the CC watermarked distribution is also i.i.d. distributed according $\tilde{Q} =  Q_{X|S}$.
On Bob's end, the detection probability is given by the expression
$$
\rd = \frac{1}{2}\left(1+ \TV\left( (PQ)^{\otimes n}, (P\tilde{Q})^{\otimes n}\right) \right),
$$
where $P\tilde{Q}(S,X)=P(S)\tilde{Q}(X|S)$
To that end, we focus on obtaining bounds on the aforementioned TV term. For a pair of distributions $P,Q$, we have the following Hellinger bounds on the TV distance  \cite{polyanskiy2024information}:
\begin{equation}\label{eq:hellinger_bounds}
    \frac{1}{2}H^2(P,Q) \leq \TV(P,Q) \leq H(P,Q)\sqrt{1-\frac{1}{4}H^2\left(P,Q\right)},
\end{equation}
where, for two measures $P,Q$ on a finite alphabet $\cX$, the squarred Hellinger divergence is given by the following equivalent forms
$$
H^2(P,Q)\triangleq \EE_Q\left[\left(1 - \sqrt{\frac{P}{Q}}\right)^2\right]=\sum_{x\in\cX}\left(\sqrt{P(x)}-\sqrt{Q(x)}\right)^2=2-2\sum_{x\in\cX}\sqrt{P(x)Q(x)}.
$$
For a pair of product distributions $(P^{\otimes n},Q^{\otimes n})$, the squarred Hellinger divergence benefits from the relation \cite{polyanskiy2024information}
$$
H^2\left(P^{\otimes n},Q^{\otimes n}\right) = 2-\left(1-\frac{1}{2}H^2(P,Q)\right)^n.
$$
Our problem therefore boils down to characterize $H^2\left(PQ,P\tilde{Q}\right)$.
We have
\begin{align*}
    H^2\left(PQ,P\tilde{Q}\right) &= \sum_{x,s}P(s)\left(\sqrt{Q(x)}-\sqrt{\tilde{Q}(x|s)}\right)^2\\
    &= \EE_S\left[H^2(Q(X),Q(X|S))\right].
\end{align*}
For a given $s,b^m)$,we have
\begin{align*}
    H^2(Q(X),Q(X|S=s) &= 2 - 2\sum_{x}\sqrt{Q(x)\tilde{Q}(x|s)}\\
    &= 2 - 2\sum_{x}Q(x)\sqrt{\frac{P_{S| Y}(s| y(x,b^m))}{P(s)}}\\
    &= 2\EE_{X}\left[1 - \sqrt{\frac{P_{S| Y}(s| Y(X,b^m))}{P(s)}}\right],
\end{align*}
where $P(S| Y)$ is the correlated channel.
Assuming $S\sim\mathsf{Ber}\left(\frac{1}{2}\right)$, we have
\begin{align*}
    H^2\left(PQ,P\tilde{Q}\right) &= 2\EE_{S,X}\left[1-\sqrt{\frac{P_{S| Y}(S| Y(X,b^m))}{P(S)}}\right]\\
    &= \EE_{ Y}\left[ 1 -\sqrt{2P(0| Y)}\right] + \EE_{ Y}\left[1 -\sqrt{2P(1| Y)}\right]\\
    &= 2 - \sqrt{2}\EE_{ Y}\left[P(0| Y)+P(1| Y)\right]\\
    &= 2- \sqrt{2}\left(\tilde{p}_0\left(\sqrt{p(0|0)}+\sqrt{p(1|0)}\right) + \tilde{p}_1\left(\sqrt{p(0|1)}+\sqrt{p(1|1)}\right) \right),
\end{align*}
where $ Y\sim\mathsf{Ber}(\tilde{p}_0,\tilde{p}_1)$.
Due to the symmetry of the correlated channel, we have for $\tilde{p}\triangleq\min(\tilde{p}_0,\tilde{p}_1)$
$$
H^2\left(PQ,P\tilde{Q}\right) = 2-\sqrt{2}f(\tilde{p})
$$
where
$$
f(\tilde{p})\triangleq \tilde{p} + \sqrt{\frac{1-\tilde{p}}{2}}\left(1 + \sqrt{1-2\tilde{p}}\right),
$$
which implies that
$$
H^2\left(P^{\otimes n},Q^{\otimes n}\right) = 2 - 2^{1-\frac{n}{2}}\left(f(\tilde{p})\right)^n.
$$
The bounds on the detection probability then follow by plugging the squarred Hellinger distance into \eqref{eq:hellinger_bounds}.$\hfill\square$

\section{Additional Information on the Numerical Solution}\label{appdx:additional_gurubi}
In this section, we outline the reformulation of the detection-perception optimization problem from the main text \eqref{eq:curve_opt_iid}, and discuss the resulting solution.
We focus on the case where $\gamma=1$, under which $\mathsf{E}_\gamma\left(Q_{X|S}, Q_X|P_S\right)=\TV\left(Q_{X|S}, Q_X|P_S\right)$.
Recall that the original optimization problem is given by
\begin{equation}
    \sup_{ Q_{X|S}}{\TV\left(Q_{X|S}, Q_X|P_S\right)},\quad \textrm{s.t.}\quad  \TV\left( \bar{Q}_X,Q_{X}\right)\leq \alpha_p.
\end{equation}

We define the optimization variable $\rV =\left[v_1,\dots,v_{k}\right]^T\in\RR^{k\times m}$ whose rows are given by $v_i = Q_{X|S=s_i}-Q_X$.
The resulting optimization problem w.r.t $\rV$ can be written as
\begin{equation}\label{eq:tv_opt_v_form}
    \begin{array}{rl}
    \displaystyle \sup_{\rV\in\RR^{k\times m}} & \|\rV\|_{1,1} \\[1.5ex]
    \textrm{s.t.} & -p\mathbf{1}_{k,m} \preceq \rV \preceq (1-p)\mathbf{1}_{k,m}, \\[1ex]
                 & \rV\mathbf{1}_{m,1} = \mathbf{0}_{k}, \\[1ex]
                 & \frac{1}{k}\left\|\rV^{\intercal}\mathbf{1}_{k,1}\right\|_1 \leq \alpha_p.
    \end{array}
\end{equation}
where $\|\rV\|_{1,1}\triangleq\sum_{i=1}^{k}\sum_{j=1}^m |v_{i,j}|$ is the entrywise $l1$-norm of $\rV$, $\mathbf{1}_{m,n}$ is an $(m,n)$ matrix of ones, and $\mathbf{0}_{m}$ is an $(m,1)$ vector of zeros.
Note that we can reduce the objective function from $\TV\left( Q_{X|S},Q_{X}|P_S\right)$ to $\|\rV\|_{1,1}$ since $P_S$ is a uniform distribution.
The first two constraints ensure that each $Q_{X|S=s_i}$ is a valid probability distribution, and the last constraint enforces $\alpha_p-$level perception. 
The resulting optimization can be further simplified. By introducing an auxiliary variable we can linearize the $\ell_1$ constraint on $\rV$. The auxiliary variable is $\rZ$ with similar dimensions to $\rV$. The resulting optimization problem is
\begin{equation}\label{eq:tv_opt_v_form_z}
    \begin{array}{rl}
    \displaystyle \sup_{\rZ,\rV\in\RR^{k\times m}} & \|\rV\|_{1,1} \\[1.5ex]
    \textrm{s.t.} & -p\mathbf{1}_{k,m} \preceq \rV \preceq (1-p)\mathbf{1}_{k,m}, \\[1ex]
    & -\rV \preceq \rZ,\quad \rV \preceq \rZ,\quad \mathbf{0}_{k\times m} \preceq \rZ \\[1ex]
                 & \rV\mathbf{1}_{m,1} = \mathbf{0}_{k}, \\[1ex]
                 & \frac{1}{k}\left\|\rV^{\intercal}\mathbf{1}_{k,1}\right\|_1 \leq \alpha_p.
    \end{array}
\end{equation}
To stabilize the optimization \eqref{eq:tv_opt_v_form_z} we introduce an additional constraint that caps the values of $\rZ$ when they exceed a certain value $M$ and are governed through an auxiliary binary matrix, which is jointly optimized with $(\rV,\rZ)$.
The optimization can be then solved using an exact solver, e.g. GUROBI \cite{gurobi}. The code implementation of this optimization problem is provided in the project's GitHub repository.

\begin{figure}[!t]
    \centering
    \includegraphics[width=0.5\linewidth]{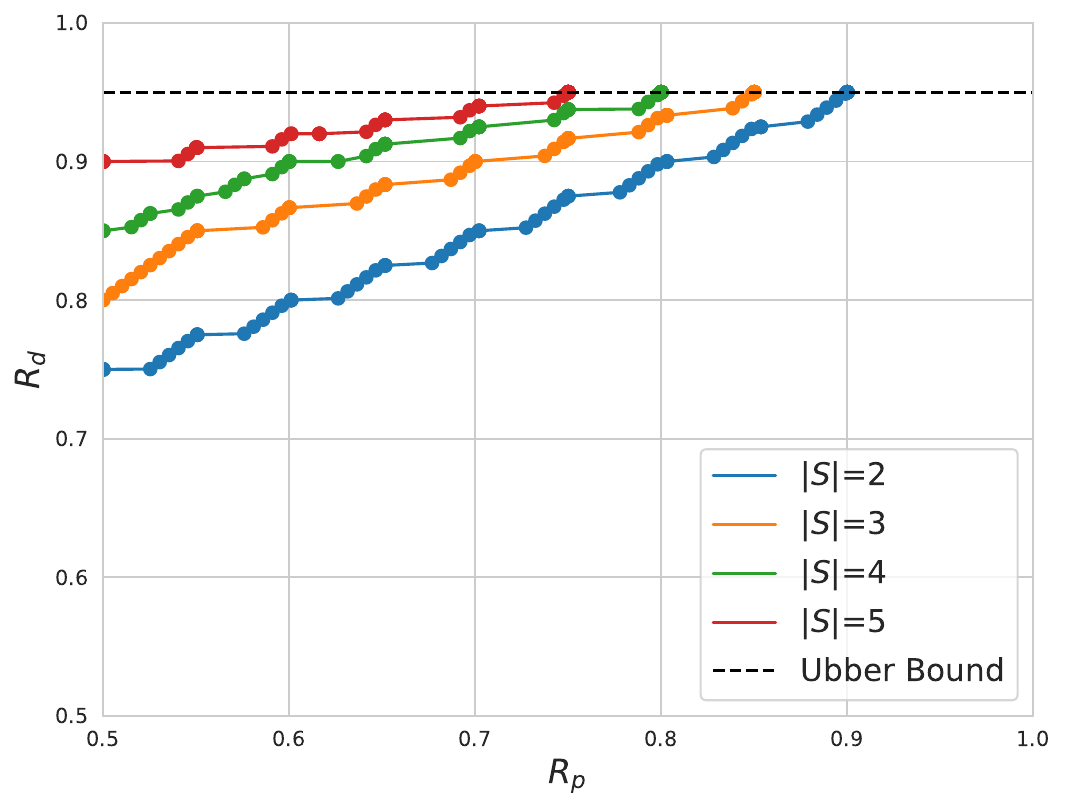}
    \caption{Numerical solution of \eqref{eq:tv_opt_v_form_z} vs. $|\cS|=k$.}
    \label{fig:tv_opt}
\end{figure}

We visualize the solution o the optimization for several values of $|\cS|=k$. The visualization is given in Figure \ref{fig:tv_opt}.
The figure depicts the optimal trade-off between detection and perception under a uniform source token distribution.
We visualize the optimization problem under an equivalence constraint for perception $(\TV(Q_x,\bar{Q}_X)=\alpha_p)$ and show the resulting curve, which is the boundary of the trade-off region. The visualization of the resulting trade-off region clearly delineated the non-convexity of the feasible set, which yields a non-convex region (below the optimality curve).
Furthermore, it can be seen that the upper bound is attained by the coupling that maximizes \eqref{fig:tv_opt}. Unfortunately, as noted in the main text, solving this optimization is infeasible in practice, due to the solver complexity and the assumption of logit access at the detector end.

\end{document}